\documentclass{article}
\usepackage[utf8]{inputenc}
\usepackage{geometry} 
\usepackage{graphicx}
\usepackage{xcolor}
\usepackage{caption}
\usepackage{subcaption}
\usepackage{graphicx}
\usepackage{bm}
\usepackage{amsmath}
\usepackage[numbers]{natbib}

\usepackage[affil-it]{authblk} 

\author[1]{Anna Guseva\thanks{anna.guseva@upc.edu}}
\author[2]{Calum Skene}
\author[2]{Steve Tobias}

\affil[1]{Department of Fluid Mechanics, Polytechnic University of Catalonia, Spain}
\affil[2]{School of Physics and Astronomy, University of Edinburgh, United Kingdom}

\graphicspath{{./}{images/}}

\title{Data-driven discovery of dynamo cycle equations}

\begin{document}
\maketitle

\begin{abstract}
    Many low-mass stars like the Sun host periodic, oscillatory magnetic fields that lead to variable levels of stellar activity, driving space weather that affects the habitability and detection of exoplanets. Owing to the intrinsic difficulty in modeling stellar magnetohydrodynamics across scales, realistic numerical simulations of this process are very challenging, and developing reduced-order models is of interest. In this work, we develop a framework to recover such models directly from numerical data by using a combination of Hankel Dynamic Mode Decomposition (DMD) to identify coherent magnetic structures, and the Sparse Identification of Nonlinear Dynamics (SINDy) framework to model their dynamics. We compare these models to those obtained using the classic mathematical method of weakly nonlinear (WNL) analysis. This approach is implemented on a one-dimensional mean-field dynamo model that parameterizes the main components of a convective dynamo in a low-mass star -- helical convection and differential rotation. Using coefficient filtering, and a constrained SINDy library, we recover oscillatory dynamo models as a function of the dynamo strength parameter $D\sim \alpha \Omega'$, magnetic dissipation parameter $\kappa$, and a comprehensive dynamo model that predicts the magnetic state for a combination of these two parameters. Our results suggest that equations discovered with SINDy are more robust than equations from WNL analysis, and can predict the saturation amplitude of magnetic fields in parameter regimes far from the onset of dynamo action characterized by stiff nonlinearities. This includes unstable, and typically unknown, subcritical branches. Further to this, SINDy is able to find equations in parameter regimes where the nonlinearity is not analytic and WNL analysis cannot be applied. These properties of data-driven SINDy models suggest them as a viable alternative for modeling of stellar dynamo cycles directly from the data.  
\end{abstract}

\section{Introduction}\label{sec:intro}

The magnetic fields of astrophysical objects often exhibit variability, including in planets like Earth and Jupiter, and in the Sun and solar-like stars with convective envelopes. In the Sun, the magnetic field varies periodically, intensifying modes of dipolar or quadrupolar parity over its 11-year cycle \citep{Tobias_Weiss_2007}. Recent spectropolarimetric and photometric data of other solar-like stars reveals magnetic cycles there as well~\citep{jeffers2023stellar}, which can affect habitability and detection of exoplanets orbiting these stars~\citep{rauer2025plato}. Although some properties of cycles are well-established, e.g. a decrease in stellar magnetic activity with stellar age and rotation slow-down, their mechanisms are still debated \citep{charbonneau2023evolution}. Both deep-seated interior effects, such as convective motions and shearing of field lines by radiative-convective interfaces, as well as surface effects of meridional advection of magnetic flux, and magnetic instabilities in the near-surface shear layer~\citep{Vasil_2024}, can be at play.  Together, these mechanisms contribute to dynamo action -- systematic generation of magnetic fields by stellar plasma flows. 

\citet{parker1955hydromagnetic} and~\citet{yoshimura1975solar} showed that dynamo cycles can be modeled as an $\alpha$-$\Omega$ dynamo, with a linear instability of magnetic field around a background flow with parametrized helical convective motions ($\alpha$-effect) and differential rotation ($\Omega$-effect). Subsequently, direct numerical simulations (DNS) of convective envelopes, were also able to recover some properties of stellar cycles, such as saturation of their period with the rotation period of stars as they age~\citep{strugarek2017reconciling}. Whilst able to explain some features of DNS dynamo cycles, such as oscillation frequencies~\citep{schrinner2011oscillatory}, $\alpha$-$\Omega$ models lack the full treatment of non-axisymmetric structures and nonlinearities in DNS and need appropriate nonlinear closures to achieve saturated, steady-state magnetic fields. The closures could include the quenching of convective $\alpha$-effect by magnetic fields, expulsion of magnetic field lines from dynamo-active region by magnetic buoyancy, or additional equations of velocity perturbations generated by the Lorentz force. Both $\alpha$-$\Omega$ and DNS models provide a far-from realistic representation of stellar magnetic fields: the first one due to the arbitrary parameterization of 3D flow effects, e.g. $\alpha$, and the latter due to their unrealistic parameter regimes caused enormous time scale separation among different stellar physics phenomena, from millions of years for magnetic field diffusion to minutes for pressure waves in stars~\citep{tobias2021turbulent}. Hence, alternative approaches to dynamo modeling are highly desired, especially those reducing the problem complexity of computationally expensive DNS.

One way to reduce the complexity is to simplify to the essential components of the dynamo -- e.g. dipoles, quadrupoles, differential rotation -- in a phenomenological dynamical model consistent with the symmetries of the full system (see e.g. \cite{1997A&A...322.1007T}). This can either be DNS or a model such as $\alpha$-$\Omega$. In each case, we can then seek parameter regimes replicating desired dynamical behavior. Such a model should ideally replicate dipolar-quadrupolar oscillations in dynamo experiments~\citep{petrelis2008chaotic}, capture some of the essential features of the solar cycle~\citep{knobloch1998modulation}, and of Tayler-Spruit dynamo based on purely magnetic instabilities~\citep{daniel2023subcritical}. Alternatively, a reduced-order model can be derived mathematically with WNL analysis, which expands the solution locally to the bifurcation point~\citep{Malkus_Veronis_1958, Stuart_1960, Watson_1960}. Strictly speaking, this is valid only near the bifurcation point, and it is challenging to apply WNL analysis to dynamos which that arise in very complex flows far away from it. 

In recent years, data-driven analysis show considerable progress in developing efficient, reduced-order models for large scales in complex fluid flows. From Galerkin projection onto Principal Orthogonal Decomposition modes~\citep{noack2011galerkin} to neural networks~\citep{bakarji2023discovering}, these methods can represent flows with just a few variables. Recently, the SINDy framework, a method for equation \textit{discovery} has attracted attention. SINDy fits the dynamics of the system onto an arbitrary library of physically plausible terms, with the weights of the corresponding coefficients reflecting the importance of each term. Terms with smaller weights are truncated to obtain a sparse model with as few terms as necessary to model the flow dynamics. Since the seminal paper of~\cite{brunton2016discovering}, SINDy has been explored for problems including cavity flow instabilities~\citep{callaham2022role}, wakes formed by blunt bodies~\citep{callaham2022empirical}, convection in ducts~\citep{loiseau2020data}, Poiseuille flow in channel flows~\citep{khoo2022sparse}, and recovery of bifurcations in discrete maps~\citep{bramburger2021data}. Promising as it seems, SINDy-identified equations can be challenging to converge for noisy systems with many degrees of freedom, and so alternative methods, such as weak (integral) formulations, have been explored~\citep{nicolaou2023data}. One of the main challenges is to establish a rigorous connection between the unknown underlying structure of a mathematically exact reduced-order model and the many possible models proposed by SINDy. In previous works, statistical approaches have been used to tackle these problems: ensemble methods~\citep{fasel2022ensemble}, or Bayesian versions of SINDy~\citep{gao2023convergence}, which can place error bars on different terms. However, a rigorous relation between mathematically and data-motivated models remains an open question. 

The aim of this work is two-fold. First, to develop robust data-driven models for canonical bifurcations to oscillatory, cyclic dynamos, using the techniques of model discovery, and second, to establish a rigorous link between these reduced-order models and WNL analysis. We will do this for a 1D $\alpha-\Omega$ dynamo model proposed by Bushby~\citep{bushby2003strong}. This model for axisymmetric fields allows the dynamo to saturate through the interaction of velocity $u$, toroidal magnetic field $B$ and poloidal potential $A$, with realistic quadratic terms akin to the induction and Lorentz forces. Since strong magnetic fields can suppress turbulent dissipation, in this work we modify this model to
\begin{eqnarray}
& A_t&  = \frac{C_\alpha \cos(\pi x /2) B}{1+\alpha_0B^2} + {\left((1-\kappa_1)+\frac{\kappa_1}{1+\kappa_2B^2}\right)}(A_{xx}  + C_\eta A), \nonumber\\
& B_t& = - D C_\Omega \sin(\pi x/2) A_x + D\left(C_1 u_x A - C_2 u A_x\right) + {\left((1-\kappa_1)+\frac{\kappa_1}{1+\kappa_2B^2}\right)}(B_{xx}  + C_\eta B), \label{eq:dynamo} \\
& u_t&  = C_3 B_x A  -  C_4 A_x B + \tau u_{xx} + \tau C_\eta^u  u \nonumber,
\end{eqnarray}
introducing the additional terms $\kappa_1$ and $\kappa_2$, controlling magnetic diffusion, as well as $\alpha_0$, providing quenching of the $\alpha$-effect, on the right-hand-side of~\eqref{eq:dynamo}. In this way, we simultaneously control both the importance of magnetic quenching and magnetic diffusion, for increasing magnetic field. Feedback of magnetic field on magnetic diffusion  introduces subcriticality to this problem, which can result in weaker subcritical and strong-field supercritical dynamos, with the field confined to regions of strong shear~\citep{tobias1996diffusivity}. Physically, this quenching can be interpreted as modification of the small-scale, non-coherent turbulent flow motions by the magnetic field, entering our model as parameters. On the other hand, the large-scale velocity field $u$ can be interpreted as magnetic modification of large-scale differential rotation, for example, torsional oscillations~\cite{covas2005dynamo}.   The dynamo number, $D = \Omega' \alpha /\eta^2 L^2$, proportional to differential rotation $\Omega'$ and the $\alpha$-effect, is the parameter driving the dynamo instability. The magnetic Prandtl number, controlling the time scale of the field and induced flow, is set here to $\tau=1$ for simplicity. In~\cite{bushby2003strong}, the constant coefficients $C_i$ are derived by integration in radius of a more realistic 2D mean-field model by assuming $\omega$-effect peaking at the bottom of the convection zone, and $\alpha$-effect in the middle of the convection zone. Yet the resulting 1D model can cover a wider range of magnetic dynamics, independently of these assumptions. $C_\alpha$ controls the strength of the convective helical $\alpha$-effect, $C_\Omega$ strength of the differential rotation, $C_{1,2}$ the Lorentz force, $C_{3,4}$ equivalent induction, and $C_\eta$, $C_\eta^u$ the diffusion of magnetic field and induced velocity, respectively. 
Throughout this manuscript, for simplicity we set $\alpha_0$ to zero as it does not change the dynamo bifurcation from supercritical to subcritical. We note that our analysis can be easily extended to the case of $\alpha_0 \ne 0$  which would allow the threshold between supercritical and subcritical behavior to be varied.

The aim of this work is not to reproduce WNL analysis exactly; as we will show later, it is challenging for dynamos with ``stiff'' nonlinearities such as those considered in \eqref{eq:dynamo}. Rather, we aim to identify how well ad-hoc SINDy models, derived with minimal constraints on the sought reduced order model, are able to reproduce the dynamics governing the data. This paper is structured as follows. In \S~\ref{sec:wnl_analysis}, we present the WNL approach and dynamo states resulting from the solution of \eqref{eq:dynamo}, and in \S~\ref{sec:data_approach} we explain the data-driven methods that we use in this work --- DMD and SINDy. In \S~\ref{sec:fixed_D_kappa} we derive reduced-order models for one fixed pair of $(D, \kappa)$. \S~\ref{sec:varD} then expands this approach to the bifurcation with the dynamo number $D$, and \S~\ref{sec:varkappa} derives a model for supercritical to subcritical transition as a function of both $D$ and $\kappa$. We discuss our results in \S~\ref{sec:discussion} and conclude in \S~\ref{sec:conclusion}. 

\section{Onset of dynamo instability}\label{sec:wnl_analysis}
\subsection{Weakly nonlinear analysis}
Using WNL analysis, an amplitude equation for an unstable mode can be obtained by expanding solutions in a small parameter governing proximity to the critical value $D^c$ where the dynamo instability onsets. To this end, we first consider the stability of equations \eqref{eq:dynamo} around the steady homogeneous solution $A=0$, $B=0$, and $u=0$, which is a fixed point of (\ref{eq:dynamo}). Linearizing about this state, and assuming perturbations with time dependence $\exp(\omega t)$, $\omega = \omega_r + \mathrm{i} \omega_i$, gives an eigenvalue problem which decouples into one for the magnetic perturbations
\begin{equation}\label{eq:eigenvalue_problem_AB}
\omega \begin{pmatrix}
    A'\\
    B'
\end{pmatrix}=
\underbrace{\begin{pmatrix}
    \partial_{xx} + C_\eta & C_\alpha \cos(\pi x /2) \\
    - D C_\Omega \sin(\pi x/2)\partial_x & \partial_{xx} + C_\eta
\end{pmatrix}}_{\mathcal{L}_{A,B}}
\begin{pmatrix}
    A'\\
    B'
\end{pmatrix},
\end{equation}
and another for hydrodynamic perturbations
\begin{equation}\label{eq:eigenvalue_problem_u}
\omega u' = \underbrace{(\tau \partial_{xx} + \tau C_\eta^u)}_{\mathcal{L}_u}  u'.
\end{equation}
As these problems decouple, we only need to consider (\ref{eq:eigenvalue_problem_AB}), since the base-state is always stable to purely hydrodynamic perturbations. Hence, the principal bifurcation to the dynamo is due to instability of the magnetic field caused by the joint action of the $\alpha$- and $\Omega$-effects. This instability onsets at the critical dynamo number $D^c$ where the system is neutrally stable, i.e. where the growth rate is zero $\Re(\omega^c) = 0$ and the frequency equals the critical one, $\Im(\omega^c)=\omega_i^c$. 

WNL analysis proceeds by considering dynamo numbers close to the critical value $D=D^c(1 + \delta\epsilon^2)$, where $\delta=\pm1$ governs whether we are at dynamo numbers above or below the critical value, respectively. To do this, we expand the full state as a series expansion in $\epsilon$. At first order we have 
\begin{gather}\label{eq:q_exp}
\textbf{q}=
(A,B)^T
=
\epsilon G(t) \textbf{q}_G \exp(\mathrm{i}\omega_i^c t) + \mathcal{O}(\epsilon^3), \\
u = \mathcal{O}(\epsilon^2),
\end{gather}
where $\mathbf{q}_G$ is the solution to the eigenvalue problem (\ref{eq:eigenvalue_problem_AB}) at $D=D^c$. As we are close to criticality, we assume that the amplitude $G$ depends on the slow timescales $t_2=\epsilon^2t$ and $t_4=\epsilon^4t$. The WNL procedure, outlined in appendices \ref{sec:a_wnl_terms_3} and \ref{sec:a_wnl_terms_5}, finds an amplitude equation for $G$ by applying solvability conditions at third order and fifth order in the series expansion.

At third order, the amplitude equation can be shown to take the form
\begin{equation}\label{eq:3rd_WNL_main}
\frac{\textrm{d} G}{\textrm{d} t} =\epsilon^2\left[\eta_3 G - \chi_3 |G|^2G\right],
\end{equation}
where the coefficients $\chi_3=\chi_u+\kappa\chi_\kappa$ and $\eta_3$, separated through their $\kappa = \kappa_1 \kappa_2$ dependence, are determined via solvability conditions.
This is the normal form of a Hopf bifurcation, with the balance between coefficients $\chi$ and $\eta$ yielding the radius of the limit cycle and the amplitude of the magnetic field resulting from the bifurcation. The sign of $\Re(\chi)$, which depends on $\kappa$, defines the type of the bifurcation. Our calculation of the coefficients demonstrate that if $\kappa <0.0017$, $\Re(\chi)>0$ and the bifurcation is supercritical: the cubic term stabilizes and saturates the exponential growth of the instability. Conversely, if $\kappa >0.0017$, $\Re(\chi)<0$ and the bifurcation is subcritical: the cubic term promotes instability and the system grows to a higher amplitude state. Physically, this is to be expected. Increasing $\kappa$ means that magnetic field growth further lowers the diffusion of the system through nonlinear effects. This is a destabilizing effect whose relative importance depends on $\kappa$.

In the subcritical case, the dynamo eventually saturates through higher-order terms. This behavior is described through the amplitude equation found at fifth order
\begin{equation}\label{eq:wnl_fifth_order}
    \frac{\textrm{d} G}{\textrm{d} t} = \epsilon^2\left[( \eta_3 + \epsilon^2 \eta_5 ) G - (\chi_3 + \epsilon^2 \chi_5 ) |G|^2 G + \epsilon^2 \mu_5 |G|^4 G\right],
\end{equation}
with the coefficients of the model being
\begin{eqnarray}\label{eq:wnl_chi35mu5}
    \chi_3 &=& \chi_{3,0}  + \kappa \chi_{3,\kappa}, \quad  \chi_5 = \chi_{5,0}  + \kappa \chi_{5,\kappa}  ,  \\
    \mu_5 &=& \mu_{5,0} + \kappa \mu_{5,\kappa}  + \kappa^2  \mu_{5,\kappa^2} + \kappa \kappa_2 \mu_{5,\kappa \kappa_2} . \nonumber
\end{eqnarray}
A detailed derivation of this equation is given in appendices~\ref{sec:a_wnl_terms_3} and ~\ref{sec:a_wnl_terms_5}. We note that, while the third-order normal form is relatively straightforward to obtain, obtaining the fifth order form is much more involved due to the large number of nonlinear interactions that take place at higher orders. 

In this work, we would like a completely data-driven approach that is able to obtain a model analogous to that obtained analytically with WNL analysis. Such a model should not rely on a preliminary linear stability analysis that involves knowledge of the overall governing equations. Hence, we need a way to obtain a timeseries for $G$ entirely from data that then leads to model discovery. However, as $G$ evolves on the slow time variables $t_2$ and $t_4$, it is not a straightforward quantity to obtain from the data without knowledge of the eigenvector, the distance from the bifurcation point $\epsilon$, and the critical frequency $\omega_i^c$. To alleviate this, we will consider the dynamics of the quantity
\begin{equation}\label{eq:H_def}
H = \epsilon G \exp(\mathrm{i}\omega_i^c t),
\end{equation}
which depends on the fast time $t$. An analytic equation for $H$ can be obtained by substituting it into~\eqref{eq:3rd_WNL_main}, yielding
\begin{equation}\label{eq:H_normal_form}
    \dot{H} = H(\mathrm{i}\omega_i^c  + \epsilon^2\eta_3) -\chi_3 |H|^2H = \eta H - \chi |H|^2H,
\end{equation}
where $\eta$ and $\chi$ are complex, and $\dot{H}$ denotes derivation with respect to the fast time $t$. A data-driven procedure for obtaining data for $H$ will be described in the next section. The fifth-order model~\eqref{eq:wnl_fifth_order} introduces a quintic nonlinearity that stabilizes the subcritical instability. In terms of the variable $H$, the fifth-order amplitude equation reads
\begin{equation}\label{eq:wnl_H_fifth}
    \dot{H} = (\mathrm{i} \omega_i^c + \epsilon^2[\eta_3 + \epsilon^2 \eta_5 ] ) H - (\chi_3 + \epsilon^2 \chi_5 ) |H|^2 H + \mu_5 |H|^4 H = \eta H - \chi |H|^2 H  + \mu |H|^4 H .
 \end{equation}

All the numerical calculations needed for this paper -- including eigenvalue problems, linear boundary value problems, and initial value problems -- are performed using the spectral PDE solver Dedalus~\citep{burns2020dedalus} using a Chebyshev discretization with a a grid-based resolution of $N=256$ points. Dealiasing is performed by computing nonlinear terms on a grid twice this size. In the model~\eqref{eq:H_normal_form}, the destabilizing term is linear and the cubic term can be either stabilizing or destabilizing. We summarize all the coefficients of the WNL model, calculated numerically, in Table~\ref{tab:wnl_coefs}. Furthermore, in Cartesian coordinates where $x = r \cos(\theta) = H_r$ and $y=r \sin(\theta) = H_i$, equation~\eqref{eq:wnl_H_fifth} reads
\begin{align}\label{eq:cart_norm_form}
    x_t &= \eta_r x - \eta_i y - \chi_r x^3 + \chi_i x^2 y - \chi_r x y^2 + \chi_i y^3  +\mu_r x^5  - \mu_i x^4 y  + 2 \mu_r x^3 y^2 - 2 \mu_i x^2 y^3 + \mu_r x y^4 - \mu_i y^5 \nonumber, \\
y_t &= \eta_i x + \eta_r y - \chi_i x^3 -\chi_r x^2 y - \chi_i x y^2 -\chi_r y^3   +\mu_i x^5 +\mu_r x^4 y + 2 \mu_i x^3 y^2 + 2\mu_r x^2 y^3 + \mu_i x y^4 +\mu_r  y^5.
\end{align}
As $H$ is complex valued, it can be represented in the polar form $H = r \exp(i \theta)$, where $r$ is the radius of the cycle in the saturated state. The radius $r$ indicates the strength of the magnetic field, and $\theta$ the phase of the dynamo cycle. Equivalent equations in polar form are
\begin{equation}\label{eq:H_polar_normal_form}
  r_t = \epsilon^2 \eta_r r - \chi_r r^3 + \mu_r r^5, \quad \theta_t = (\omega_i^c + \epsilon^2 \eta_i) -  \chi_i r^2 + \mu_i r^4.
\end{equation}
Thus, reduced-order modeling of the dynamo system can be performed both in polar and Cartesian form; in this work, we will compare performance of the two approaches. 
\setlength{\tabcolsep}{3pt}
\begin{table}[t]
    \centering
    \begin{tabular}{ccccccc}
        \hline 
        \multicolumn{7}{l}{Critical dynamo number and frequency:} \\\hline
        &$D^c$ &  $\omega_i^c$ &  && &\\
        &$260.12$  & $-25.53$& & & &\\ \hline \hline
       $\eta$: &$\eta_3$ &  & $\eta_5$ &&&\\
        &$11.20 - 17.09i$ & &$-1.36+2.33 i$ &&&\\\hline
      $\chi$:   &$\chi_{3,0}$ &$\chi_{3,\kappa}$  &$\chi_{5,0}$& $\chi_{5,\kappa}$ && \\ 
         &$ 0.058 + 0.011 i$ & $-33.96+11.64 i$ & $0.046 - 0.0096 i $& $ -40.28 +14.01 i$ &&\\  \hline
      $\mu$:    &&&$\mu_{5,0}$&$\mu_{5,\kappa}$ & $\mu_{5,\kappa \kappa}$ & $\mu_{5,\kappa \kappa^2}$\\
          & &&$-6.1 \cdot 10^{-5}+0.00022 i$&$-0.20 - 0.088 i$ & $122.27-4.12 i$ & $-100.54+36.95 i $\\ \hline
    \end{tabular}
    \caption{Critical parameters at the onset of instability and the WNL terms of the fifth-order normal form as given by equations~\eqref{eq:wnl_chi35mu5}, \eqref{eq:wnl_H_fifth}.}
    \label{tab:wnl_coefs}
\end{table}

\subsection{Dynamo states from nonlinear simulations}
\begin{figure}[t]
     \centering
     \begin{subfigure}[b]{0.47\textwidth}
         \centering
         \includegraphics[width=\textwidth]{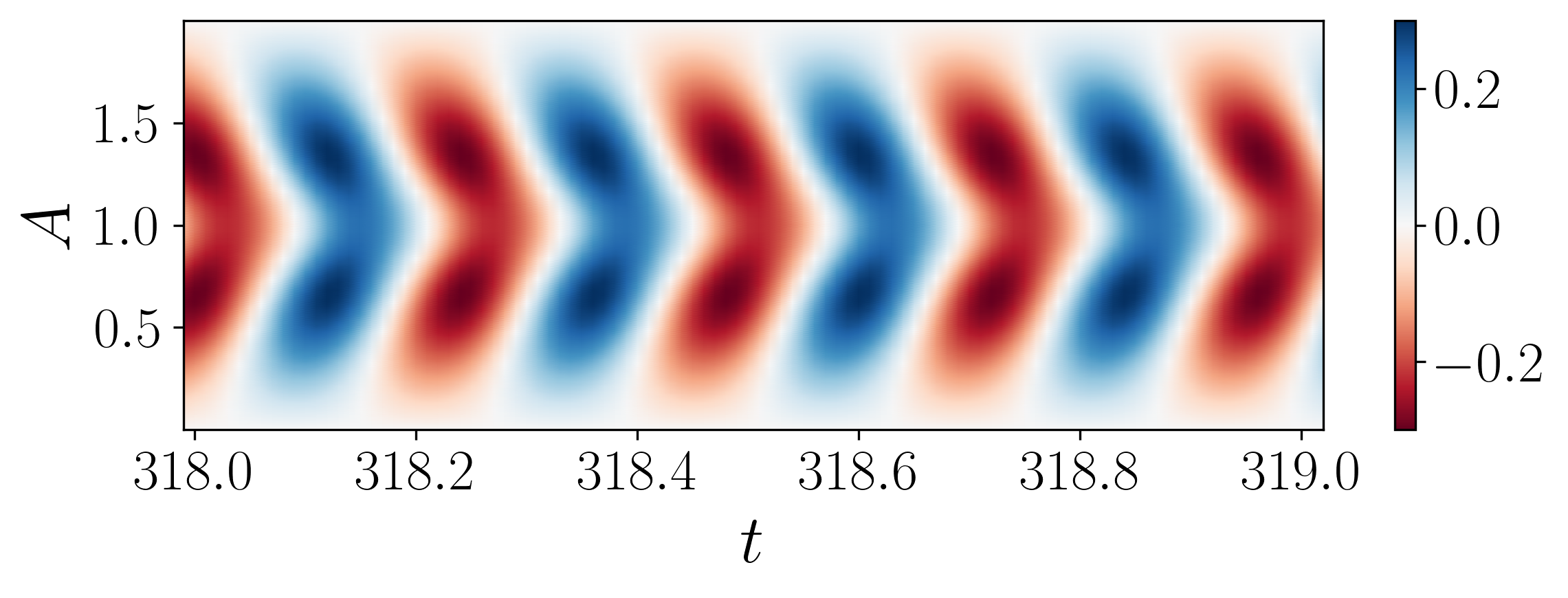}
          \caption{}
     \end{subfigure}
     \hfill
     \begin{subfigure}[b]{0.47\textwidth}
         \centering
         \includegraphics[width=\textwidth]{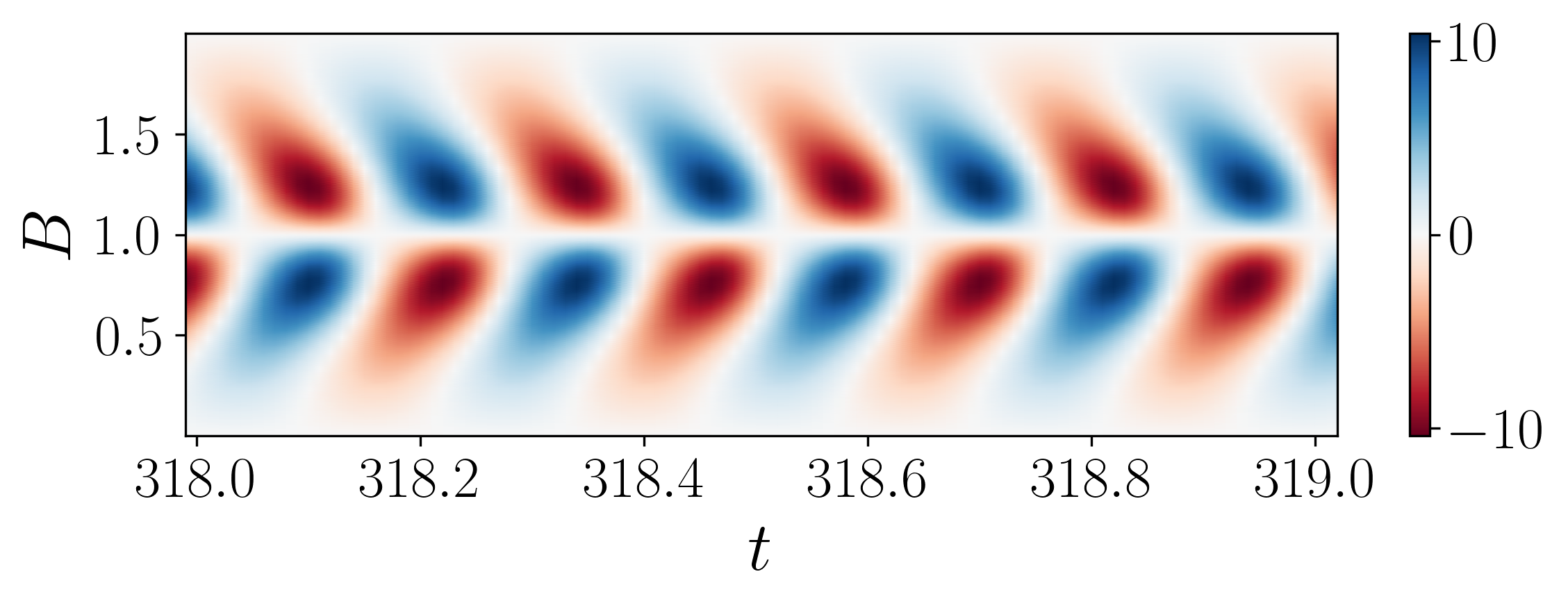}
          \caption{}
     \end{subfigure}
     \vfill
     \begin{subfigure}[b]{0.45\textwidth}
         \centering
         \includegraphics[width=\textwidth]{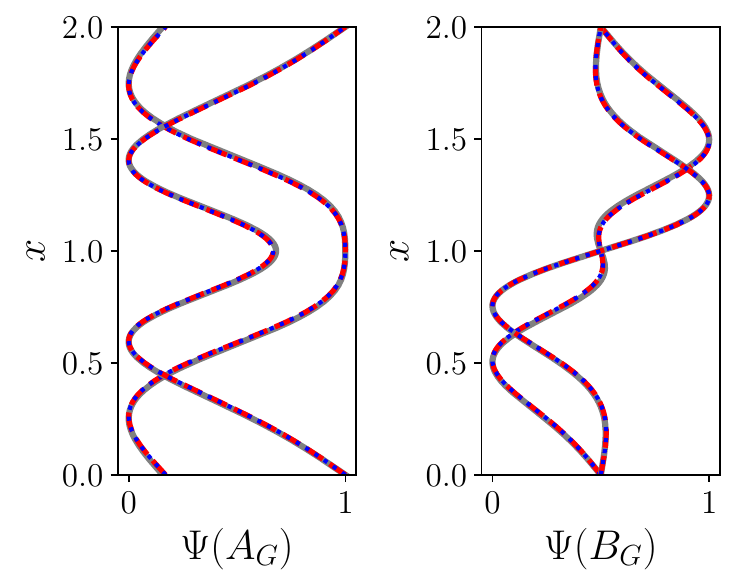}
          \caption{}
     \end{subfigure}
     \hfill
          \begin{subfigure}[b]{0.45\textwidth}
         \centering
         \includegraphics[width=\textwidth]{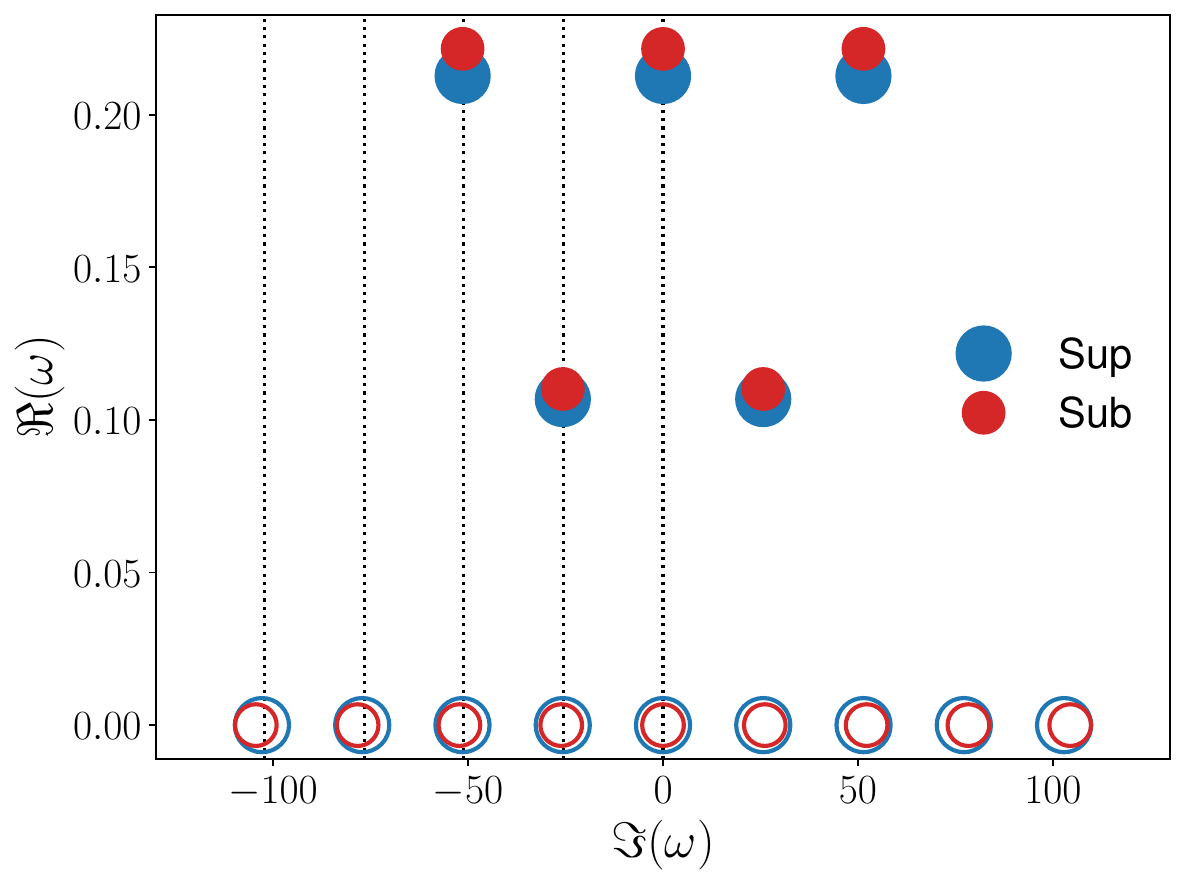}
         \caption{}
     \end{subfigure}
        \caption{Dynamo topology at the onset of instability, as a function of latitude $x$ and time $t$. (a) Magnetic potential $A$; (b) Toroidal field $B$. Subcritical dynamo with $\kappa_1=0.4$, $\kappa_2 = 0.005$, $D=262.725$ ($\epsilon = 0.1$).  (c) Real and imaginary part of the eigenvector (solid grey); real and imaginary part of the first DMD mode in the steady state. Red dashed line, supercritical case; blue dotted line, subcritical case. During the linear regime, DMD modes are identical to the eigenmodes. (d) DMD spectrum of the dynamo for supercritical case (in blue) and subcritical case (in red). Full symbols, DMD of the linear dynamo phase; empty symbols, DMD of saturated state.}
        \label{fig:solutions_dmd}
\end{figure}

To obtain temporal data of the dynamo in the form of snapshots for subsequent data-driven model reduction, we solve the nonlinear dynamo equations (\ref{eq:dynamo}) numerically using an implicit-explicit third-order four-stage Runge--Kutta time-stepping scheme. The simulations are initialized with a small perturbation taking the spatial structure of the eigenvector, to exclude any additional degrees of freedom that could complicate the dynamics of~\eqref{eq:H_normal_form} or~\eqref{eq:wnl_fifth_order}. We set the length of the system to $L=2$, and the coefficients of~\eqref{eq:dynamo} to $C_\alpha=C_\Omega = 2.39$, $C_\eta = -9.0$, $C^u_\eta = -18.23$, $C_1 = -1.29$, $C_2 = 0.29$, $C_3 = -1.02$, $C_4 = 0.79$, similarly to~\citep{bushby2003strong}. The rest of the parameters of the model are set as discussed in section~\ref{sec:intro}. For this parameter set, dynamo instability occurs when $D>D^c\approx260$ as given by the linear stability analysis (table~\ref{tab:wnl_coefs}). For both subcritical and supercritical dynamos, the dynamo instability arises in the form of an anti-symmetric magnetic structure which propagates towards the equator (figure~\ref{fig:solutions_dmd}), similarly to the solar magnetic cycle. The toroidal field $B$ is anti-symmetric with respect to the equator (figure~\ref{fig:solutions_dmd}b), while the poloidal magnetic potential $A$ is symmetric(figure~\ref{fig:solutions_dmd}a) . The period of oscillations, $T\approx 0.225$, is in agreement with the critical frequency $\omega_i^c$ (table~\ref{tab:wnl_coefs}). With the same magnetic topology, subcritical and supercritical dynamos have different saturation mechanisms and intensity for the same value of $D$.

\section{Data-driven reduced-order modeling}\label{sec:data_approach}

\subsection{Dynamic Mode Decomposition}\label{sec:dmd_method}
We extract the timeseries for $H$ from the dynamo snapshots using DMD to identify periodic components of dynamical systems~\citep{schmid2022dynamic}. DMD seeks a linear operator $\mathcal{A}$ that maps the flow state at time $t_k$ to the flow state at time $t_{k+1}$, i.e. $\bm{q}_{k+1} = \mathcal{A} \bm{q}_k$. To improve the quality of decomposition, we use high-order, or Hankel variation of the DMD method \citep{le2017higher}, as it better separates the contributions of different frequencies in modal coefficients~\citep{guseva2024data}. We construct an augmented data matrix from the flow snapshots, i.e.
\begin{equation}\label{eq:HankelQ}
Q = 
\left[
\begin{array}{cccc}
    \bm{q}_1 & \bm{q}_2 & \cdots & \bm{q}_{K-d + 1} \\
    \bm{q}_2 & \bm{q}_3 &  \cdots  &  \bm{q}_{K-d + 2} \\
     \vdots &  \vdots & & \vdots  \\
    \bm{q}_{d} & \bm{q}_{d+1} &  \cdots & \bm{q}_{K} \\
\end{array} \right],
\qquad Q' \approx \mathcal{A} Q. 
\end{equation}
Truncating the SVD decomposition of the data vector data vector $ Q = \Phi \Sigma V^* \approx \Phi_r \Sigma_r V_r^*$ at $r$ significant components, or degrees of freedom, we search for eigenvalues  $\lambda$ and eigenvectors $\tilde{\psi}$ of reduced matrix $\mathcal{A}_r$ in the vector space of the leading principal orthogonal components of $Q$:

 \begin{equation}\label{eq:Ar}
  Q' \approx  \mathcal{A} \Phi_r \Sigma_r V_r^*, \qquad \Phi_r^* \mathcal{A} \Phi_r = \Phi_r^* Q' V_r \Sigma^{-1}_r = \mathcal{A}_r.
\end{equation}
Here $d$, the so-called delay parameter, and model rank $r$ are \textit{ad-hoc} parameters of Hankel DMD that are defined by the user.
The corresponding DMD modes $\psi$ and DMD eigenvalues $\omega = \Re(\omega)+\mathrm{i} \Im(\omega)$ can be found as
 \begin{equation}\label{eq:DMD_modes}
 \psi = \frac{1}{\lambda} Q' V_r \Sigma^{-1}_r \tilde{\psi}, \qquad \omega = \ln(\lambda)/ \Delta t.
 \end{equation}
At given $\omega$, the components of $\psi$ should resemble the components of expansions~\eqref{eq:AppA_exp} and~\eqref{eq:expansion_u}, e.g. $\mathbf{q}_G$, $u_{G^2}$, $u_{|G|^2}$. 
To obtain $H$ as defined in~\eqref{eq:H_def}, we need to estimate how much weight each component of $\psi$ has in the data, i.e. we must identify the modal temporal coefficients of expansion  
 \begin{equation}
 Q \approx \sum c_i (t) \psi,
 \end{equation}
and, in particular, the coefficient corresponding to the ``eigenvalue" mode $\psi^G$, $c^G$.
 
Unlike the Principal Orthogonal Decomposition modes $\Phi_r$, DMD modes are not orthogonal with respect to themselves. Projecting the data $Q$ on them can be therefore done only with oblique projection, for example, least-squares best fit of the data on the modes. This inevitably leads to the loss of information or mixing of temporal dynamics from different modes, especially of those with similar spatial structure. To address this problem, we develop an adjoint-based method to calculate the coefficients of the DMD modes. After obtaining the ``direct" DMD modes and eigenvalues from~\eqref{eq:DMD_modes}, we calculate the eigenmodes $\psi^\dagger$ of the conjugate-transpose, adjoint matrix $\mathcal{A}^\dagger_r$. The eigenmodes of this matrix are orthogonal with respect to the DMD modes. Furthermore, by normalizing the direct eigenmodes with the vector product $<\psi^\dagger, \psi>$ we obtain a set of orthonormal components that satisfy $\psi \psi^\dagger = I$, where $I$ is the identity matrix. Now, the temporal coefficients of the direct DMD modes can be obtained simply by multiplication of the data vector $Q$ by the corresponding adjoint mode with a complex-conjugate eigenvalue, since
\begin{equation}
c_i (t) = Q \psi_i^\dagger = c_i \psi_i \psi_i^\dagger = c_i(t).
\end{equation}
This gives us exact information about the amplitude of each mode at a fixed instance in time. An alternative exact method would be to project the data on adjoint eigenvector $\mathbf{q_G^\dagger}$ which gives us the value of $H$ directly. However, such an approach is not entirely data-driven.

\subsection{DMD analysis of the data}

During the development of the dynamo instability, system~\eqref{eq:dynamo} transitions from the linear regime to a fully saturated state. Since the DMD algorithm does not require a statistically steady state, it can be performed on the data from either of these two phases of flow evolution, with each regime revealing the physics of growth and saturation, respectively. One of the key parameters of DMD is its rank $r$. For our model, only one pair of complex-conjugate modes, corresponding to the dominant eigenvectors of the dynamo wave, is sufficient to contain $99$\% of the magnetic energy. Nevertheless, this rank is too restrictive to separate these eigenvectors from the contribution of higher order, nonlinear, interactions suggested by the WNL expansion (e.g.~\eqref{eq:AppB_exp3}). We thus increase the DMD rank $r$ to recover the corresponding flow components.

In the linear regime ($t<50$), DMD with rank $r=5$ allows us to recover the two complex-conjugate dominant eigenvalues with frequencies $\pm \omega_i^c$, the mean flow with $\Im(\omega)=0$ driven by them, and the second-order quadratic flow components with the eigenvalue of $\pm 2 \omega_i^c$ (\ref{fig:solutions_dmd}d). The real part the $\pm \omega_i^c$ modes is consistent with the growth rate obtained from a linear stability analysis at $D = 262.725$, and the growth rates of the secondary modes are two times faster than those of the eigenvectors; which is expected since they are formed from quadratic interactions of the fundamental mode. The data from the saturated periodic states of the dynamo were sampled on an interval of the same length, $250<t<300$, and number of time steps. Unlike the exponential growth phase, modes obtained from periodic data are neutrally stable and have nearly zero growth rates  (empty symbols in figure~\ref{fig:solutions_dmd}d). When increasing the rank to $r=9$, this DMD recovers cubic and quadratic frequencies as well, corresponding to higher-order nonlinearities in~\eqref{eq:AppA_exp},~\eqref{eq:expansion_u}. This indicates that higher-order nonlinearities become stronger in the saturated state, as expected. It is possible to recover even higher frequencies by increasing $r$. We found  $r=9$ to be a reasonable DMD basis size to recover the dynamics of eigenvectors in the steady state. These results hold for supercritical and subcritical bifurcations, with similar DMD eigenvalue spectrum in both cases (figure~\ref{fig:solutions_dmd}d).

\begin{figure}[t]
    \centering
     \begin{subfigure}[b]{0.47\textwidth}
         \centering
         \includegraphics[width=\textwidth]{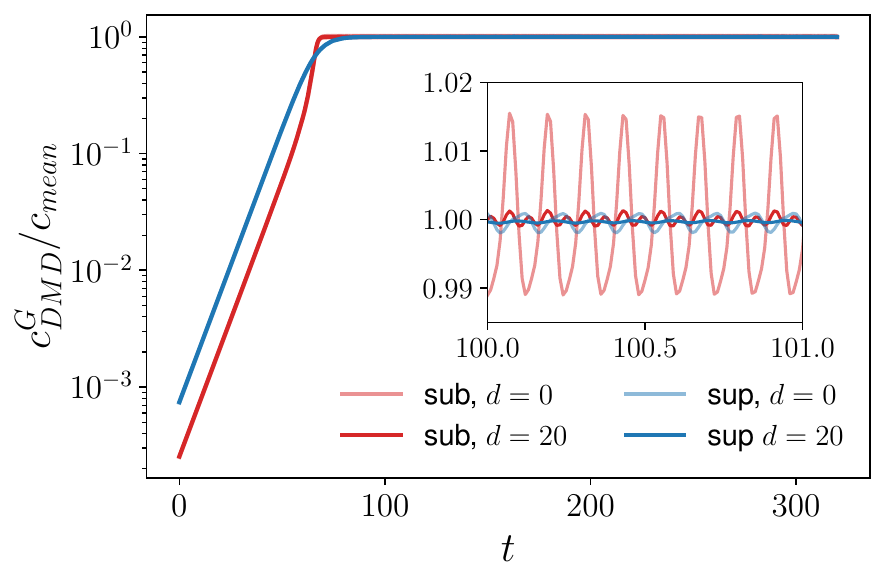}
          \caption{}
     \end{subfigure}
     \hfill
         \centering
     \begin{subfigure}[b]{0.47\textwidth}
         \centering
         \includegraphics[width=\textwidth]{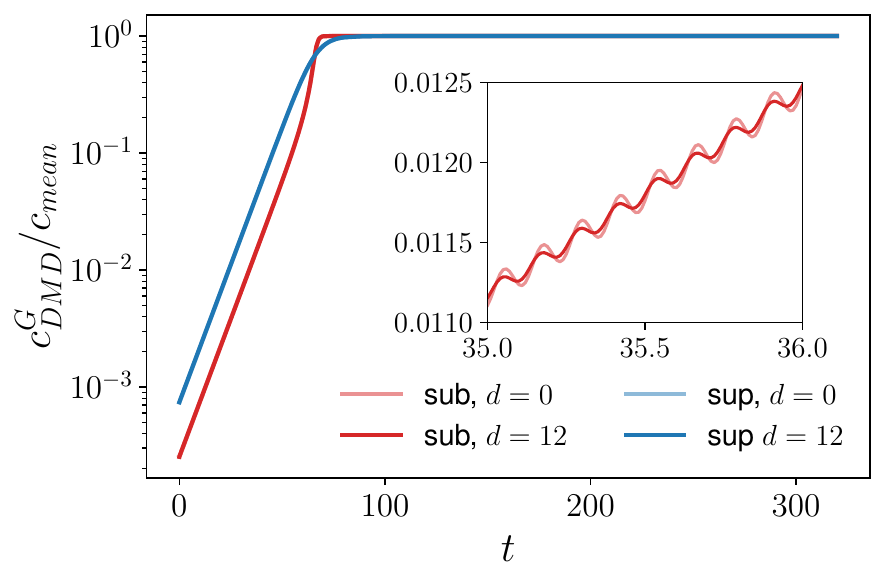}
          \caption{}
     \end{subfigure}

    \caption{ Temporal coefficient of the leading DMD mode calculated with adjoint DMD. (a) DMD on the linear evolution phase, $t<50$, $r=5$; (b) DMD on steady-state data, $t>270$, $r=9$. Blue, subcritical dynamo bifurcation; red, supercritical dynamo bifurcation. $\epsilon = 0.1$; $D = 262.725$.}
    \label{fig:adj_coef}
\end{figure}

The real and imaginary components of the DMD modes in the linear phase are identical to the corresponding components of the system eigenvector (figure~\ref{fig:solutions_dmd}c). These mode pairs respect the primary symmetry of the data, symmetric for $A$ and anti-symmetric for $B$ about the equator, and can be viewed as a pair of oscillators with the corresponding eigenfrequency~\citep{guseva2024data}. The temporal coefficients for the evolution of these modes are obtained by taking a product with adjoint DMD modes, as described in section~\ref{sec:dmd_method}, and are in a good agreement with the values of $H$ extracted from simulations using an inner product of the data with the adjoint eigenvector (giving the exact projection onto the neutral eigenvector which WNL analysis provides an amplitude equation for). The absolute values of these coefficients reflects the temporal evolution of the eigenvector amplitude (in blue, figure~\ref{fig:adj_coef}), with the exponential growth  nonlinearly quenched at about $t\approx 60$, for this initial condition. The subcritical case exhibits a steeper exponential growth around this time, due to the destabilizing cubic term in the normal form~\eqref{eq:wnl_H_fifth}, and is then abruptly quenched by higher-order terms (in red, figure~\ref{fig:adj_coef}). In both cases, the steady-state value of $H$ corresponds to the radius of the resulting limit cycle of the Hopf bifurcation. As $D$ increases, the DMD modes begin to deviate from the eigenvectors.

Panels (a) and (b) in figure~\ref{fig:adj_coef} highlight the difficulties of obtaining the temporal coefficients of the eigenvectors from the data. In figure~\ref{fig:adj_coef}(a), the DMD coefficients, extracted during the linear regime, develop oscillations in the nonlinear phase. Similarly, the DMD modes that were extracted in the saturated regime feature oscillations in the linear phase in figure~\ref{fig:adj_coef}(b). In fact, the exact values of $H$ from simulations also develops similar oscillations (see figure~\ref{fig:wnl_sindy}a). We interpret these oscillations not as DMD artifacts but the contributions from second-order quadratic terms (as their frequency is approximately $2 \omega_i^c$). They are potentially dangerous for system identification with SINDy due to its sensitivity to noisy derivatives~\citep{brunton2016discovering}; in our case, the data and corresponding derivatives are smooth so such oscillations invariably lead to overfitting of the normal form. To remove these oscillations, we use high-order DMD, i.e. we increase the delay $d$ and which yields better separation of components with different frequencies~\citep{guseva2024data}. The inset plots in figure~\ref{fig:adj_coef} show that increasing the delay considerably reduces the amplitude of the oscillations in the modal coefficient. Varying the delay parameter $d$ for the subcritical and supercritical bifurcations, at fixed rank $r=5$ and $r=9$, allows us to identify the optimal decomposition delay $d$ that minimizes the amplitude of the oscillations. This delay was found to be  $d=18-20$ for the linear basis with rank $r=5$ and $d=12$ for the steady-state basis with rank $r=9$. We remark that, it is difficult to define \textit{a priori} whether the linear or steady-state basis gives the most robust results for the modeling; the linear phase gives the closest decomposition to the actual eigenvectors and so allows us to better reproduce initial state, whereas the steady-state basis reduces quadratic oscillations during the dynamo saturation which is important in recovering higher-order nonlinearities.

\subsection{Sparse Identification of Nonlinear Dynamics (SINDy)}\label{sec:sindy_method}
DMD, described above, reduces the number of degrees of freedom in the system by representing the whole data set as a product of modes $\psi_i$ and temporal coefficients $a_i (t)$, $Q \approx \sum_i a_i(t)$. Despite the linearity assumption of DMD, these coefficients are nonlinearly correlated (e.g., see~\citep{callaham2022role}). WNL forms of our dynamo model are sparse equations, containing select nonlinear terms. In this work, we will test the capability of  SINDy~\citep{brunton2016discovering}, to recover them. The SINDy algorithm is general, and allows us to recover both ordinary and partial differential equations with arbitrary libraries of terms. In this work, we perform SINDy regression in the reduced space of the flow represented by $a_i(t)$ and their temporal derivatives $\dot{a}_i = d a_i / d t$. From that, we construct a set of possible linear and nonlinear functions $F_j (a_i)$, forming a library of potentially important terms that describe $\dot{a}_i$. The choice of functions $F(a_i)$ depends on the physical problem; a polynomial set of functions $a_i^j$, $j=0,1,2,...$ is a reasonable choice for systems with quadratic linearities and triadic interactions, and will be used here. By minimizing the $L^2$ norm between derivative and the functions,
\begin{equation}\label{eq:regressionSINDy}
    ||\dot{a}_i - \Theta F(a_i)||_2,
\end{equation}
a first guess on the coefficient magnitudes is obtained. Then, the coefficients $\Theta$ smaller than a certain threshold $\Theta_{cr}$ are disregarded and set to zero. The regression is then performed again, without taking into account previously eliminated terms, for a fixed number of iterations ($N=10^4$ in our case). This results in sparse model where many terms in $\Theta$ are zero. The resulting reduced model is then integrated in time, allowing for both the final error in the approximation of the right-hand-side (RHS), $\textrm{error}^{\textrm{RHS}}$, and the error in the dynamics, $\textrm{error}^{\textrm{LHS}}$, to be evaluated:
\begin{equation}\label{eq:errorSINDy}
   \textrm{error}^{\textrm{RHS}} = ||\dot{a}^{\textrm{DNS}}_i - \Theta F(a_i)||_2, \qquad   \textrm{error}^{\textrm{LHS}} = ||\dot{a}^{\textrm{model}}_i - \dot{a}^{\textrm{DNS}}_i||_2.
\end{equation}
These errors are put together in Pareto plots to help choose the optimal threshold $\Theta_{cr}$ that allows for the sparsest model with the least number of terms and with the smallest error. When varied, the dynamo number $D$ and damping $\kappa$ are introduced as additional variables in the regression~\eqref{eq:regressionSINDy}. In these cases, the error~\eqref{eq:errorSINDy} is evaluated with respect to the highest values of $D$ and $\kappa$, to optimize the models for dynamos further away from the onset of linear instability.

Since our aim is to reconstruct normal forms such as~\eqref{eq:H_normal_form} and~\eqref{eq:wnl_H_fifth}, we will apply SINDy regression only on the first two leading DMD modes corresponding to the eigenvectors of the dynamo system. At first, we will reconstruct the normal forms for fixed parameters $D$ and $\kappa$; then, we will obtain normal forms as a function of $D$ where $\kappa$ is fixed, and lastly, we will move on to constructing a normal form reflecting both $D$ and $\kappa$ dependence. To perform SINDy regression, we use open-source Python package PySINDy~\citep{Kaptanoglu2022}. We note, that after obtaining a model for the leading DMD modes, we could train a further model that obtains higher order mode behavior directly from the dynamics of this pair~\citep{callaham2022role}. This would allow us to reconstruct the entire flow data~\citep{callaham2022role}. However, we do not do this here as we concentrate on comparison with normal forms.

\section{Supercritical and subcritical normal forms at fixed $D$ and $\kappa$}\label{sec:fixed_D_kappa}
\subsection{Data preparation and analysis}\label{sec:data_prep}

Due to the presence of oscillations from quadratic interactions, SINDy regression is prone to overfitting, i.e. the resulting models are more complex than the desired normal forms. To avoid this, and to obtain robust sparse models resembling WNL normal forms, we prepare the data as follows.
\begin{enumerate}
    \item First, the data are collected from simulations with the same parameters as in figure~\ref{fig:solutions_dmd}, i.e. $\epsilon=0.1$  ($D = D^c (1+ \epsilon^2) = 262.725$), $\kappa_1 = 0$ and $0.4$ ($\kappa_2= 0.005$), with a time lag of $\Delta t = 0.01$. Simulations are initialized with the flow eigenvectors to remove initial transients where stable directions decay. To increase the data sample seen by SINDy, the amplitude of the initial condition is randomly varied, with the data collected from $N=9$ different simulations. 
    \item The data vectors $A$, $B$ and $u$ are normalized with their standard deviation, and are stacked together in a single array. Without normalization, the magnitude of $B$ is larger than the magnitude of $A$, and the components of DMD coefficients does not match the eigenvectors. The velocity can be omitted since its contribution to the eigenvector is exactly zero at the first order, since the instability is purely magnetic. However, we found that including it in the data vector helps to separate quadratic contributions, as it gets projected onto the ``second-order" DMD modes with $\pm 2\omega_i^c$.
    \item Hankel DMD is performed on the data vector retaining up to $5$ modes for the linear regime basis and up to $9$ modes for the steady-state basis. This reduces oscillations yet does not remove them completely (see figure~\ref{fig:adj_coef}).
    \item To obtain modal coefficients, we take the inner product of the data vector with the adjoint DMD modes. The complex coefficient $c(t) = x(t) + i y(t)$, corresponding to the dominant mode and representing the eigenvector, is retained and separated into its radius and phase,  $ c(t) = r(t) \exp(\mathrm{i}\phi(t)) $.
    \item To remove the remaining contribution of nonlinear interactions to the coefficient, the coefficients are filtered with a Savitsky--Golay filter with a typical window length of $200-300$ points and third order polynomials. This allows us to remove oscillatory noise in the derivatives and avoid overfitting. For rapidly growing and strongly subcritical models, this filtering was performed twice with a smaller window length to avoid spurious effects during saturation and at the end of the time series. This procedure leads to smoother finite difference derivatives that enables a more robust sparse regression~\citep{de2020pysindy}.
    \item\label{norm_step} The radii $r$ are normalized with their steady-state absolute values, $r_{norm}$. This normalization brings the library terms to the same order of magnitude, and helps to avoid spurious high-order polynomial terms in the regression. In addition, the $i \omega t$ component of the phase can be subtracted, as is done in section~\ref{sec:fixed_D_kappa}, to improve the reconstruction fidelity of nonlinear terms in the phase equation. 
\end{enumerate}
This procedure leads to robust results for the SINDy, that are less sensitive to derivation methods and the truncation threshold.

\subsection{Identified normal forms}\label{sec:indentified_normal_forms}

After normalization (step \ref{norm_step} in section~\ref{sec:data_prep}), the coefficients of the corresponding normal forms are proportional to the original normal form coefficients (table~\ref{tab:wnl_coefs}) multiplied by a polynomial factor of the normal form radius for each corresponding term of~\eqref{eq:H_normal_form} or~\eqref{eq:wnl_H_fifth}. To facilitate the comparison, we re-normalize the original WNL normal forms with the corresponding radii of the limit cycle, and subtract the constant term in phase corresponding to the imaginary part of the eigenvalue. In this form, the corresponding third-order WNL equations for supercritical dynamo at $D=262.725$ ($\epsilon = 0.1$), $\kappa_1 \kappa_2=0$ read
 \begin{eqnarray}\label{eq:sup_wnl}
r_t &=& 0.112 r - 0.0589r^3, \nonumber \\
\theta_t &=& -0.171 - 0.0105 r^2.
 \end{eqnarray}
 
\begin{figure}[t]

    \centering
     \begin{subfigure}[b]{0.47\textwidth}
         \centering
         \includegraphics[width=\textwidth]{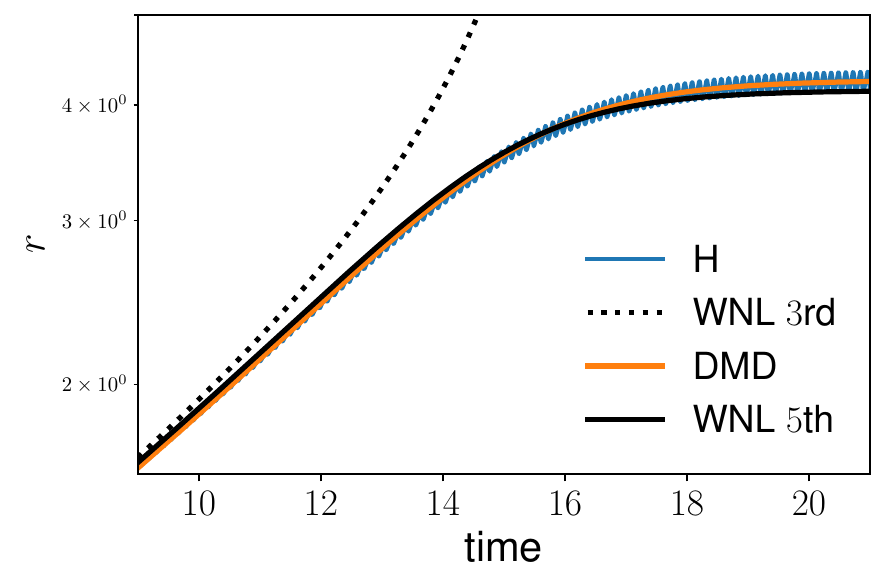}
         \caption{}
     \end{subfigure}
     \hfill
         \centering
     \begin{subfigure}[b]{0.47\textwidth}
         \centering
         \includegraphics[width=\textwidth]{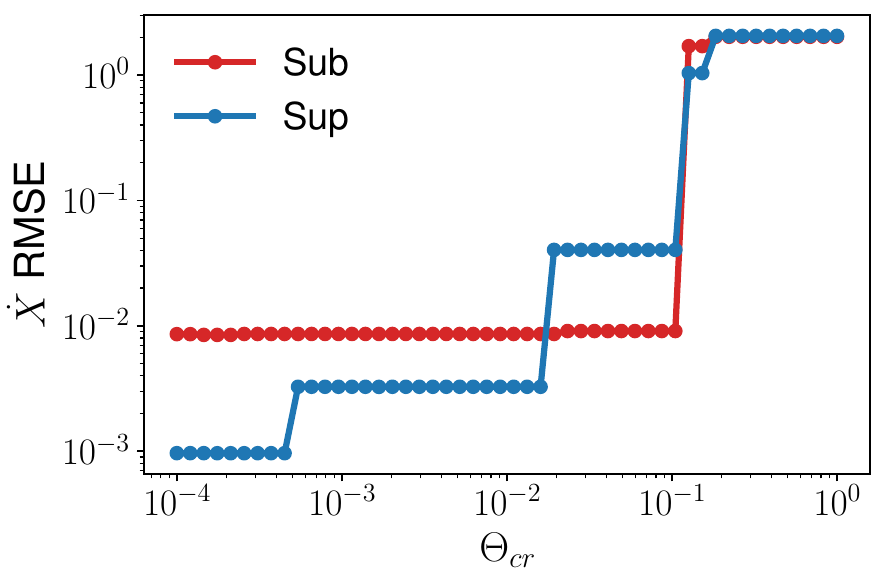}
          \caption{}
     \end{subfigure}
     \vfill
         \centering
     \begin{subfigure}[b]{0.47\textwidth}
         \centering
         \includegraphics[width=\textwidth]{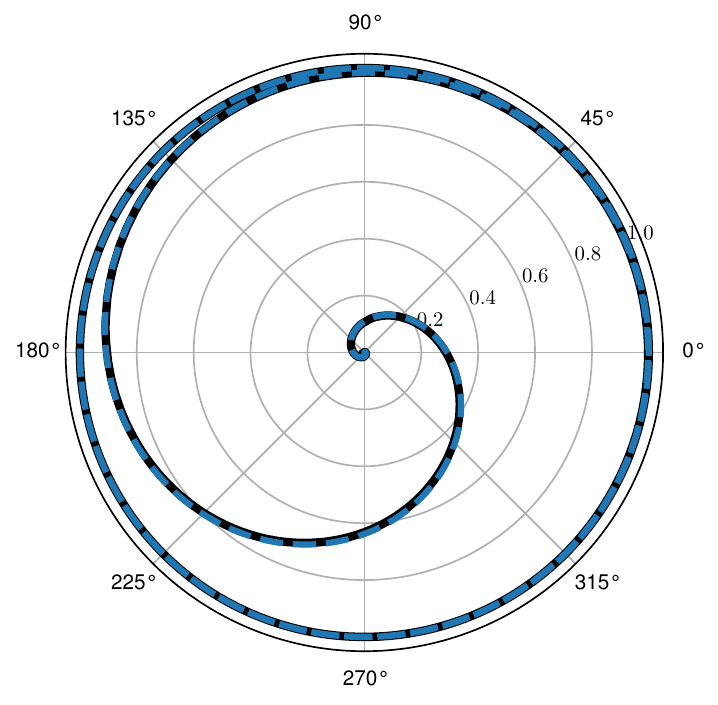}
          \caption{}
     \end{subfigure}
     \hfill
         \centering
     \begin{subfigure}[b]{0.47\textwidth}
         \centering
         \includegraphics[width=\textwidth]{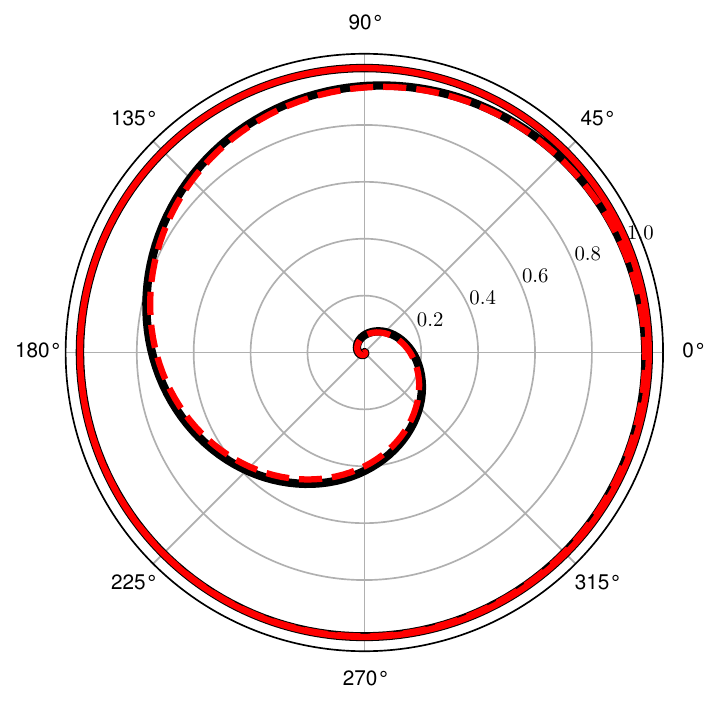}
          \caption{}
     \end{subfigure}


    \caption{(a) Comparison of eigenvector amplitude $H$, DMD modal coefficient $c_{DMD}^G$, third-order and fifth-order WNL models for subcritical case:  $\kappa_1 =0.4$, $\kappa_2 = 0.005$. (b) Relative mean squared error of the solution as a function of SINDy cut-off threshold $\Theta_{cr}$ along the entire trajectory in figure~\ref{fig:adj_coef} for supercritical case (in blue) and subcritical case (in red). (c) Temporal integration of the SINDy models (dashed), compared to the data (black): (c) supercritical equations~\eqref{eq:sup_wnl_sindy}, $\kappa=0$; (d) subcritical equations~\eqref{eq:sub_wnl5_sindy},  $\kappa_1 =0.4$, $\kappa_2 = 0.005$.  $D=262.7$ ($\epsilon=0.1$) for all panels.}
    \label{fig:wnl_sindy}
\end{figure}

To contrast, we choose a second data set with the same level of supercriticality, $D=262.725$, but in the mildly supercritical regime $\kappa_1 = 0.4$, $\kappa_2 = 0.005$ ($\kappa_1 \kappa_2 = 0.002 >0.0017$). In this case, a third-order nonlinear model is linearly unstable and its numerical solution increases exponentially until it diverges. Fifth-order damping terms are necessary to counteract the destabilizing third order term and bring the dynamo to a steady state. Similarly to before, we normalize the fifth-order WNL model~\eqref{eq:wnl_H_fifth}, with the radius of the limit cycle at these parameters, obtaining
\begin{eqnarray}\label{eq:sub_wnl5}
     r_t &=& 0.112 r + 0.175  r^3 -1.293 r^5, \nonumber\\
\theta_t &=& -0.171 - 0.612 r^2 +0.117 r^4. 
 \end{eqnarray}
 A typical temporal evolution of these models is shown in figure~\ref{fig:wnl_sindy}(a). The exact values of $|H|$, obtained using the inner product of the adjoint eigenvector $q_G^\dagger$ with the solution at each time step, contains a quadratic oscillatory component, showing that the contribution of the oscillatory components to $A$ and $B$ is not negligible in the data, unlike that given by~\eqref{eq:AppA_exp}. Model~\eqref{eq:sub_wnl5} diverges without fifth-order terms, and saturates with them. However, it saturates to a slightly higher value of $H$ than the corresponding DMD mode. 
 
 We recover models~\eqref{eq:sup_wnl} and~\eqref{eq:sub_wnl5}, by applying SINDy regression on the processed DMD coefficients, as explained in section~\ref{sec:data_prep}. To establish the optimal model with the smallest error and least number of terms, we sweep over a range of thresholds in the interval $[10^{-4}, 1]$, obtaining models with different complexity. We estimate the performance of these models, as described in section~\ref{sec:sindy_method}, and track both errors as a function of the cut-off threshold for the coefficients $\Theta_{cr}$, in the form of Pareto plots (figure~\ref{fig:wnl_sindy}b). As $\Theta_{cr}$ decreases and more terms are included in the model, the accuracy increases and the error manifests abrupt decrease. When incorporating further terms of higher order we obtain insignificant improvements in accuracy and plateau of the error, a signature of overfitting and non-optimal SINDy models. The minima in Pareto plots guide our choice for $\Theta_{cr}$ in this and the following sections. In addition, here we aim for models which are as close to the WNL normal form as possible and thus always choose thresholds that lead to the full, WNL-like system of equations.

 The Pareto plot (Figure~\ref{fig:wnl_sindy}(b)) shows the error of the solution vector for supercritical and subcritical bifurcations, as discussed above. For both models, the first sharp decrease in the model error takes place at about $\Theta_{cr}=0.1$, which is sufficient to obtain equation~\ref{eq:sub_wnl5} for subcritical bifurcation with its large normalized coefficients. For the supercritical bifurcation, however, the nonlinear coefficient in the phase in~\eqref{eq:sup_wnl} is too small, and choosing $\Theta_{cr} =0.1$ eliminates the quadratic phase correction. These terms are recovered after the second abrupt decay in the Pareto plot, $\Theta_{cr} =0.015$. The corresponding SINDy model for the supercritical bifurcation, based on the adjoint coefficient obtained for steady-state evolution, is
 \begin{eqnarray}\label{eq:sup_wnl_sindy}
r_t &=&  0.108 r - 0.108 r^3, \nonumber\\
\theta_t &=& -0.166 - 0.0187 r^2,
 \end{eqnarray}
 which gives an excellent agreement with the data (figure~\ref{fig:wnl_sindy}a). SINDy regression for the subcritical bifurcation, based on the steady-state evolution basis and Hankel DMD  with $r=9$, $d=12$,  gives the following optimal model:
\begin{eqnarray}\label{eq:sub_wnl5_sindy}
     r_t &=& 0.110 r +  0.249 r^3 - 0.179 r^4 - 0.179 r^5 \nonumber, \\
\theta_t &=& -0.166 - 0.641 r^2 + 0.133 r^4.
 \end{eqnarray}
 Compared to the WNL equations~\eqref{eq:sub_wnl5}, this model has the correct structure of the phase equation, with reasonably close values of coefficients. In radius, the sign and value of the corresponding coefficients in $r$ and $r^3$ are also comparable. However, instead of the strong fifth-order damping term, SINDy fits a combination of the fourth and the fifth-order terms, with nearly equal coefficients. Least-squares regression is not able to distinguish between the two for subcritical data; this behavior does not improve if more data is incorporated or if the time interval is constrained only to the saturation interval around $t\approx 70$ (figure~\ref{fig:adj_coef}). Nevertheless, numerical integration of this model is again in excellent agreement with the data (figure~\ref{fig:wnl_sindy}d). The relative error of both models is less than $1\%$, and so they provide a good approximation near the onset of the dynamo instability. 
 \setlength{\tabcolsep}{3pt}
 \begin{table}[t]
    \centering
    \begin{tabular}{|c|c|c|c|c|c|c|c|c|c|c|c|c|c|c|c|}\hline
           Model  &\multicolumn{6}{|c|}{Equation for $r$} & \multicolumn{6}{|c|}{Equation for $\theta$} & Error\\\hline
     Polynomials &1 & $r$& $r^2$& $r^3$&$r^4$&$r^5$  &1 & $r$ & $r^2$ & $r^3$ & $r^4$ & $r^5$ &\\\hline 
    Normalization   & & $1$& & $r_{norm}^2$&&$r_{norm}^4$  &1 & & $r_{norm}^2$ &  & $r_{norm}^4$ &  &\\\hline 
      \multicolumn{14}{|c|}{Supercritical bifurcation}  \\\hline 

      WNL*, third order   &  & 0.112 & &-0.059&  & & -0.171 & & -0.0110& & & &\\\hline 
      Linear regime    & & 0.108 & & -0.108 &  & &  -0.166 & & -0.0187& & &&$0.11\%$\\\hline 
      Steady-state    & & 0.108 & &-0.108 &  & &-0.166 & & -0.0187 && &&$0.11\%$\\\hline  \hline
      \multicolumn{13}{|c|}{Subcritical bifurcation}  \\\hline
      WNL*, fifth order    && 0.112 & &0.175 &&  -1.293  & -0.171&  &-0.612 & & 0.117& &\\\hline 
      Linear regime    &  &  0.110 & & 0.245 & -0.173 & -0.182  &  -0.166&  &-0.638 & & 0.130& &$0.23\%$\\\hline 
      Steady    & &  0.110 & &  0.249 & -0.179&  -0.179 & -0.166 &  & -0.641& & 0.133 & &$0.24\%$\\\hline \hline
     Constrained SR3   & & 0.114 & & 0.159 &&-0.273  & & &  & & && 0.96\%\\ \hline \hline

            Bagging, mean   & & 0.110 & &0.245 & -0.172 & -0.182 & &  & & & &&0.26\%\\\hline 
      Bagging, std  &  &$3.5 \cdot 10^{-5}$ & & $8.6 \cdot 10^{-4}$ & $1.7 \cdot 10^{-3}$& $8.9\cdot 10^{-4}$  & & &  & & &&\\\hline 
    \end{tabular}
    \caption{Coefficients of the data-driven reduced models, compared with the WNL analysis. WNL models were normalized to facilitate comparison. Column ``1" refers to a constant ``bias" term, also included in SINDy library. The coefficients of the WNL form in table~\ref{tab:wnl_coefs} were normalized with $r_{norm} = 55.4$ and $r_{norm} = 71.4$ for supercritical and subcritical bifurcation, respectively. The last column corresponds to the error of the integrated model, compared to the actual DMD coefficient.}
    \label{tab:coefs_sindy}
\end{table}

To test the robustness of the models, table~\ref{tab:wnl_coefs} compares the results for different strategies of data collection for SINDy regressions. The third-order supercritical model is robustly recovered both from the basis derived from linear and steady-state phases of flow evolution, because nonlinearities are not strong enough. In comparison, when the data is taken from the linear or steady-state phase of the subcritical bifurcation, a spurious quadratic term appears in radius.


The equation for $r$ is independent of $\theta$ and thus we can check the presence of spurious terms in the subcritical SINDy equations for $r$ separately. To do so, we perform two additional types of SINDy regression for the $r$-equation. The first one is ensemble, or bagging SINDy~\citep{fasel2022ensemble}, which calculates an ensemble of SINDy models based on splitting of the input data set, and calculates the corresponding means and standard deviations of the coefficients. All ensemble SINDy models yield the same structure as before, with the mean values of the coefficients nearly identical to the previous models (table~\ref{tab:coefs_sindy}). The standard deviation of the coefficients is thus expectedly small, less than one percent. In a second test, we constrained the coefficients of $r^2$ and $r^4$ terms to be zero using the constrained SR3 optimizer~\citep{zheng2018unified}, instead of sparse least-squares regression. This way, regression is performed only on the nonlinearities expected in the normal form~\eqref{eq:sub_wnl5}, so only the values of the coefficients, rather than the equation structure, has to be discovered. However, the resulting equation is still differs from the WNL equations, with much weaker damping at the fifth order. The corresponding relative error of the model is also considerably larger. It is likely that the presence of the fourth order term in $r$ is due to the contribution to $A$ and $B$ from quadratic terms not captured by the WNL model. Overall, we find that regression is robust to data both from the linear and steady dynamo phases, and it is only the higher-order subcritical nonlinearities that are recovered incorrectly.
 
\section{Model as a function of dynamo number $D$}\label{sec:varD}

The dynamo number $D$, representing the joint action of helical turbulence and rotation in our models, is the main bifurcation parameter of the system. However, in real systems we sometimes may not know the distance from subcriticality, or the expansion needed to obtain the corresponding normal form (in our case, $D= (1+ \delta\epsilon^2) D^c$). In this section, we take the analysis from section~\ref{sec:fixed_D_kappa} as a starting point, and derive dynamo models as a function of $D$. After that, we will compare this direct approach with the WNL theory, far away from the onset of the dynamo instability.

\subsection{Supercritical model}\label{sec:varD_sup}
To obtain a supercritical model, we collect data for several dynamo simulations near the onset of instability with $D \in [240, 280]$. We calculate the DMD modal basis $D=265$, where the leading DMD modes still do not deviate from the eigenvectors. As previously, we perform HODMD with $r=9$ and $d=12$ on the data from the steady state. To use as little insight about the model from linear analysis as possible, we also did not subtract the eigenvalue part, $\mathrm{i} \omega_i^c t$, from the phase. The DMD modal coefficients are then formulated in terms of $r$ and $\theta$, filtered with Savitsky--Golay filter of similar width of $(300, 3)$ and normalized with the radius of the normal form at $D=265$ where the eigenmodes have been calculated. Using the optimal SINDy threshold of $\Theta_{cr}=0.02$, and adding the dynamo number $D$ as an additional variable in the SINDy library, we obtain the following reduced-order model: 
\begin{align}\label{eq:sup_varD}
    r_t &= (-11.014 + 0.042 D) r - 0.207 r^3 = \eta_r(D)  r - \chi_r r^3 \nonumber ,\\
\theta_t &= -8.273 -0.066 D - 0.026 r^2 =  \eta_i(D) - \chi_r r^2.
\end{align}

We now compare this expression with the normal forms~\eqref{eq:H_normal_form} and~\eqref{eq:sup_wnl}, and the corresponding coefficients in table~\ref{tab:coefs_sindy}, with quadratic and cubic terms that are consistent with the analytical form. The first terms in radius, $\eta_r (D) r$, and in phase, $\eta_i (D)$, both have a constant part and a part depending on the dynamo parameter $D$. This is expected from the original dynamo PDE, containing linear terms both proportional to and independent of $D$, and also from the corresponding normal form~\eqref{eq:H_normal_form}.  The negative part of the linear term can be interpreted as diffusive damping, counteracting instability at small $D$. The correction to the phase, proportional to $D$, approximates the increase in oscillatory dynamo frequency with $D$. At $D=265$, where the basis was calculated, $\eta_r (D) = 0.116$ and $\eta_i (D) = 25.733$, again consistent with the normal form~\eqref{eq:sup_wnl}. At smaller $D$, $\eta_r (D)$ decreases until it becomes negative and the instability ceases to exist. In this approximate model this occurs slightly above the exact $D^{c}$. The cubic term $\chi$ is also largely in agreement with the values from table~\ref{tab:coefs_sindy}.

\begin{figure}[t]
 \centering
     \begin{subfigure}[b]{0.47\textwidth}
         \centering
         \includegraphics[width=\textwidth]{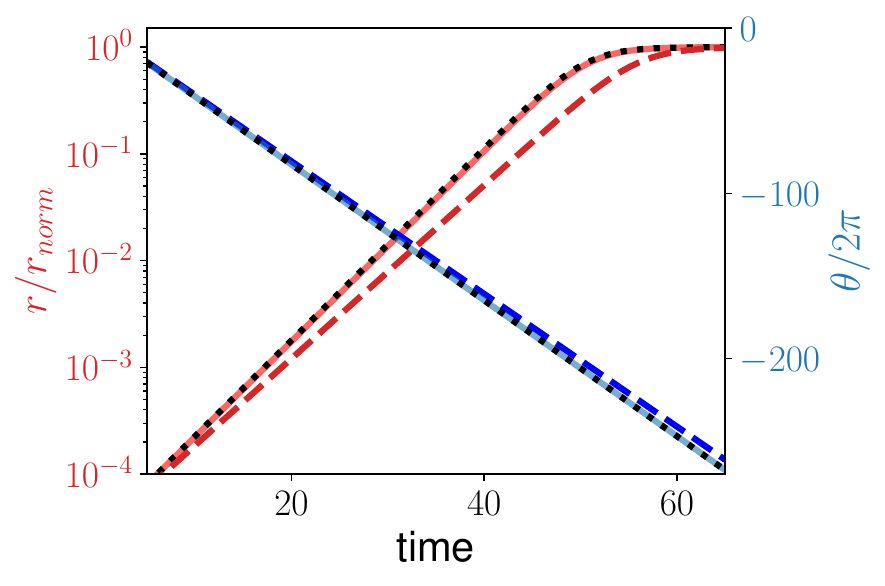}
          \caption{}
     \end{subfigure}
     \hfill 
      \centering
     \begin{subfigure}[b]{0.47\textwidth}
         \centering
         \includegraphics[width=\textwidth]{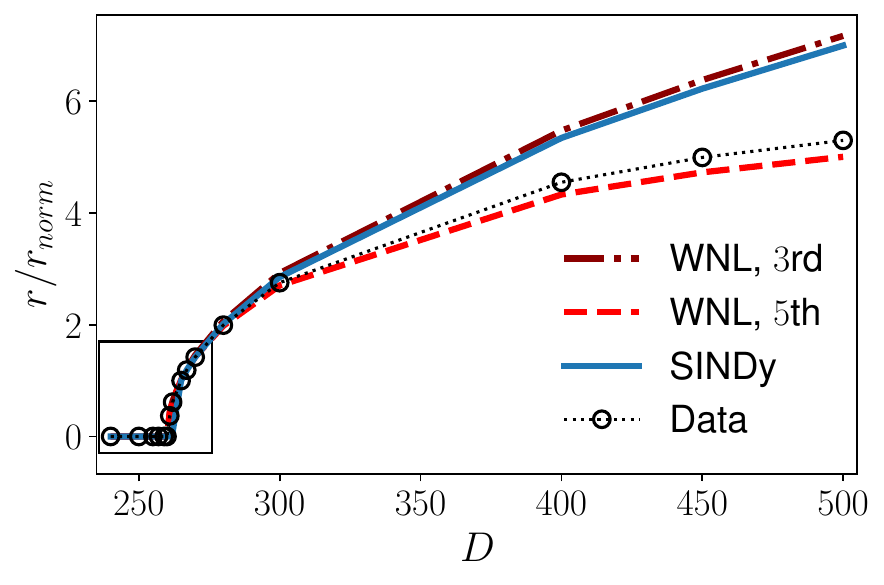}
          \caption{}
     \end{subfigure}
     \vfill
      \centering
     \begin{subfigure}[b]{0.47\textwidth}
         \centering
         \includegraphics[width=\textwidth]{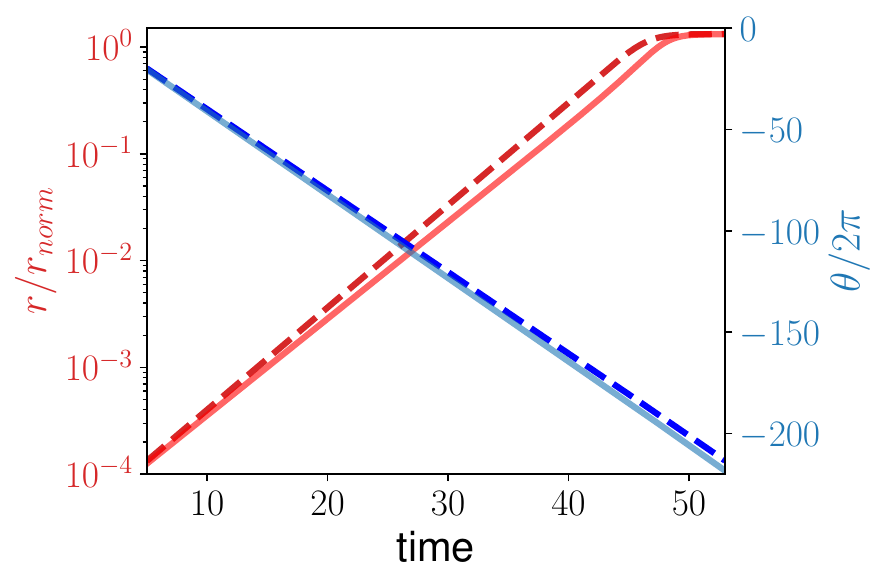}
          \caption{}
     \end{subfigure}
     \hfill
      \centering
     \begin{subfigure}[b]{0.47\textwidth}
         \centering
         \includegraphics[width=\textwidth]{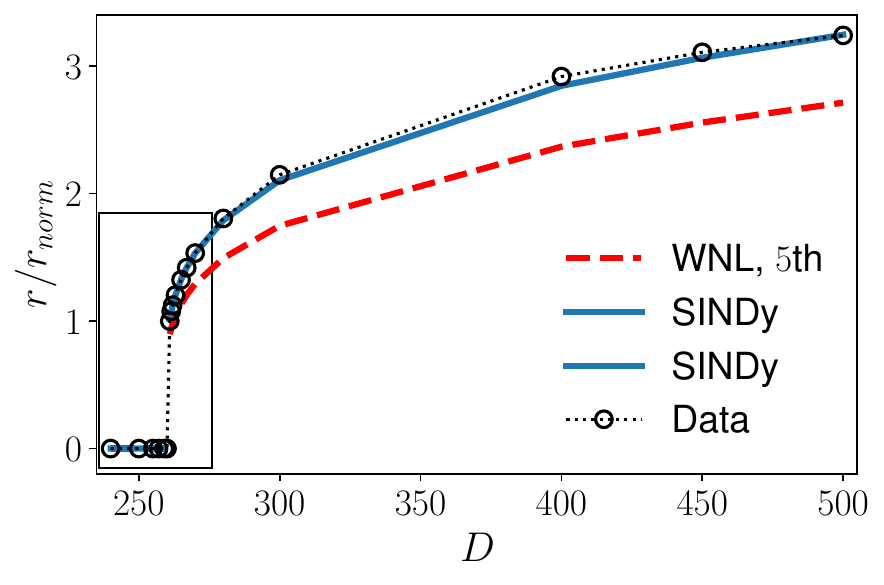}
          \caption{}
     \end{subfigure}
     
    \caption{  (a) Temporal evolution supercritical dynamo ($\kappa=0$) given by models~\eqref{eq:sup_varD} and~\eqref{eq:sup_cart_varD} with $D=265$.  Solid lines correspond to the data, dotted lines to polar model~\eqref{eq:sup_varD}, dashed to Cartesian model~\eqref{eq:sup_cart_varD}. Radius is denoted in red and phase in blue. (b) Magnitude of the dynamo cycle, normalized with the magnitude at $D=265$.  Circles, DMD coefficients from the data; dash-dotted line, third order WNL~\eqref{eq:H_normal_form}; dashed, fifth order WNL~\eqref{eq:wnl_H_fifth}; solid line, third order SINDy model~\eqref{eq:sup_cart_varD}.  (c) Same as (a) but for subcritical bifurcation, $\kappa_1 = 0.4$, $\kappa_2 = 0.005$. (d) Same as (b) but for subcritical bifurcation. Circles, DMD coefficients from the data;  dashed, fifth order WNL model~\eqref{eq:wnl_H_fifth}; solid line, fifth order SINDy model~\eqref{eq:sub_cart_varD}.}
    \label{fig:model_varD}
\end{figure}

Alternatively, a way to recover the normal form is to perform SINDy regression in the space of ``Cartesian" coordinates of the real ($x$) and imaginary ($y$) parts of the dominant DMD coefficients. Filtering the coefficient in $(r, \theta)$-space and using the optimal threshold of $\Theta_{cr} = 0.01$, the following model was obtained:
\begin{align}\label{eq:sup_cart_varD}
    x_t = (-10.648  + 0.041 D) x - (-8.939 - 0.063 D)y - 0.201 x^3 + 0.029 x^2 y - 0.200 x y^2 + 0.030 y^3, \\
y_t = (-8.872 - 0.063 D)x  + (- 10.634  + 0.041 D)y - 0.028 x^3 - 0.201 x^2 y -0.027 x y^2 - 0.200 y^3 \nonumber.
\end{align} 
It can be shown that this model is mathematically equivalent to the ``polar" normal form~\eqref{eq:sup_varD}, with a good qualitative agreement with the values of $\eta$ and $\chi$. In figure~\ref{fig:model_varD}(a) we compare the performance of this model to the polar model~\eqref{eq:sup_varD}. The ``Cartesian" model has more discrepancy in the radius during the exponential growth of the dynamo, and weaker instability growth rate than its polar version. The discrepancy in the phase of the solution compared to the data is also growing faster. Nevertheless, both models approach the same amplitude of the magnetic field in the steady-state, are robust and insensitive to perturbations in initial conditions, and converge to the same attractor. However, equations~\eqref{eq:sup_varD} are much faster to integrate in time than~\eqref{eq:sup_cart_varD}.

Both~\eqref{eq:sup_varD} and~\eqref{eq:sub_cart_varD} can be easily extrapolated up to $D=10^4$. In figure~\ref{fig:model_varD}(b) we compare the results of such an extrapolation. The figure shows limit-cycle amplitude predictions from SINDy reduced-order models, WNL predictions and the actual amplitudes from DMD, plotting them as a function of $D$. The black rectangle denotes the region where the data for SINDy models have been collected; i.e. the regression did not use any data outside this area. The WNL and both SINDy models all give a good approximation of the dynamics at the onset of instability. Away from the onset, $D>350$, the actual amplitude of dynamo waves is systematically smaller than the one predicted by both SINDy and the WNL models with only cubic nonlinearities. This leads to large errors in the magnetic field amplitude, up to $35\%$. Only the fifth-order WNL model reproduces this decrease in the dynamo amplitude, mostly due to the correction to the cubic term $\epsilon^2 \chi_{5,0}$ in~\eqref{eq:wnl_H_fifth}. Then, the error reduces to only $5\%$ with respect to the data. 

\subsection{Subcritical model}\label{sec:varD_sub}
We aim to identify a similar $D$-dependent normal form, but for the subcritical bifurcation with fixed $\kappa_1 = 0.4$ and $\kappa_2 = 0.005$, in the interval of $D\in [240, 270]$, similarly to section~\ref{sec:fixed_D_kappa}. A subcritical dynamo model requires including fifth order terms in the SINDy library. With the parameter $D$ as the third variable, the number of terms in the fifth order library increases considerably, and sparse regression fails to converge to correct solution in polar coordinates $(r, \theta)$, eliminating all the terms even for reasonably small thresholds. Constraining the library only to the polynomial terms in $r$ and $D$ leads to non-sparse models overfitting in radius and phase.
In ``Cartesian" coordinates, no constraints on the library are necessary and SINDy directly converges to the expected fifth-order equation. Regression on the real and imaginary part of the coefficients gives us the following set of equations:
\begin{align}\label{eq:sub_cart_varD}
    x_t & = (-10.99 + 0.042 D) x - (-8.449 - 0.065 D) y + 0.035 x^3 + 0.428 x^2 y + 0.034 x y^2 + 0.43 y^3\nonumber  \\ &-0.096 x^5 -0.058 x^4 y -0.192 x^3 y^2 -0.116 x^2 y^3 -0.096 x y^4 -0.058 y^5 \nonumber,\\
y_t & = (-8.391 -0.065 D) x + (-10.945 + 0.042 D) y -0.425 x^3 + 0.029 x^2 y -0.428 x y^2 + 0.03 y^3 \\ & + 0.057 x^5 -0.093 x^4 y + 0.114 x^3 y^2 -0.188 x^2 y^3 + 0.058 x y^4 -0.094 y^5 .\nonumber 
\end{align}
This model was obtained with normalization of $r$ at $D=262$ and the SINDy threshold of $\Theta_{cr}=0.1$. The linear terms in~\eqref{eq:sub_cart_varD} are similar to the supercritical model~\eqref{eq:sup_cart_varD}, and the model also possesses a high degree of symmetry between different coefficients. Comparing~\eqref{eq:sub_cart_varD} with~\eqref{eq:cart_norm_form} gives us bounds on the WNL coefficients. Near the onset, e.g. $D=265$, they read
\begin{align}
\eta_r \in [0.140, 0.185],  & \quad \eta_i \in [-25.674, -25.616], \nonumber\\
  \chi_r \in [-0.035,-0.029],  & \quad \chi_i \in [0.425, 0.43] ,\\
 \mu_r \in [ -0.096,-0.093],  & \quad \mu_i \in [0.057, 0.058] \nonumber.
\end{align}
These values are comparable with those in table~\ref{tab:coefs_sindy} and equation~\eqref{eq:sub_wnl5}, even though the third and the fifth-order terms are smaller. This is likely due to the difference in normalization of the DMD coefficients: data in model~\eqref{eq:sub_wnl5} was normalized with the radius of the limit cycle $\kappa = 0.002$, $D = 262.7$. In contrast, coefficient amplitudes in the data set for model~\eqref{eq:sub_cart_varD} were normalized so that $|r_c| \geq 1$ for  $D \geq 262$, requiring weaker damping of high-polynomial orders at large $D$ and $r>1$. Additionally, high-order coefficients in the subcritical WNL model~\eqref{eq:wnl_H_fifth} in reality depends on both distance from $D^{c}$ and the subcriticality parameters $\kappa_1$ and $\kappa_2$, which SINDy models based on $D$ do not capture. Despite these discrepancies, model~\eqref{eq:sub_cart_varD} approximates the data much better than WNL at the onset of the dynamo instability, and away from it (figure~\ref{fig:model_varD}d), extrapolating well into large $D$. The relative error of model ~\eqref{eq:sub_cart_varD} with respect to the data is about $2.5\%$ while the fifth-order WNL error is about $19\%$. Interestingly, unlike for supercritical bifurcation, this error does not increase with $D$ (figure~\ref{fig:model_varD}b). We will discuss this effect further in section~\ref{sec:discussion}.

The results of sections~\ref{sec:fixed_D_kappa} and~\ref{sec:varD} suggest that SINDy regression on ``Cartesian" libraries containing the real and imaginary parts of the DMD modal coefficients is more robust than regression on the library in polar form (the phase and radius). Although models with additional terms in radius like~\eqref{eq:sub_wnl5_sindy} approximate the local dynamics reasonably well, they can generate non-physical, for the original system, fixed points -- for example, fixed points with negative radii. In the following, we will seek only for ``Cartesian" dynamo models.

\section{Dependence on the subcriticality parameter}\label{sec:varkappa}
\subsection{Models at fixed $D$ and varying $\kappa$}\label{sec:varkappa_fixedD}

The dependence of the WNL normal form~\eqref{eq:wnl_H_fifth} on the dynamo number $D$ is relatively straightforward since it enters the linear terms at $\mathcal{O}(\epsilon^2)$ directly, and affects the growth rate of the dynamo instability. The constant eigenvalue contribution to $\eta$ is always the largest term, followed by $\epsilon^2 \eta_{3,0}$; the fifth order contribution $\epsilon^4 \eta_{5,0}$ is always smaller for the considered values of $D$.  For the cubic terms in equation~\eqref{eq:wnl_chi35mu5}, the constant value of $\chi_{3,0}$ is linearly balanced by $\kappa \chi_{3, \kappa}$ terms; their competition controls transition from subcritical to supercritical bifurcations (figure~\ref{fig:SINDy_varkappa_varD}a). The rest of the contribution from the fifth order analysis, $\epsilon^2 \chi_{5,0}$ and $\kappa \epsilon^2 \chi_{5,\kappa}$ are at least an order of magnitude smaller; terms proportional to $\kappa$ are typically very small and become large only at large $\kappa$  (see~\ref{eq:wnl_chi35mu5}). For the fifth-order terms, the contribution to $\mu$ from these terms becomes comparable to the constant part of the coefficient relatively fast.

First, we test the capability of SINDy to recover the dependence on $\kappa$ at a fixed value of the dynamo number, $D = 262$. As previously, we set $\kappa_2 = 5 \cdot 10^{-3}$, and span several values of $\kappa_1$ so that $\kappa = \kappa_1 \kappa_2 \in [0, 2.25 \cdot 10^{-3}]$, and introduce $\kappa$, $\kappa^2$ and their products with $x$ and $y$ in the library. Since $\kappa_2$ is constant, the corresponding coefficient will contribute to the coefficients of the proportional to $\kappa$ terms, and so $\kappa_2$ is not added to the library as an independent variable. The parameter $\kappa$ is very small, this brute-force regression onto the full fifth-order library results in a non-sparse model with numerous spurious products of $\kappa$, $x^2$ and $y^2$ (not shown here). The weights of the spurious coefficients are typically quite large. Essentially, we observe overfitting, as small-valued library functions, even when multiplied by a large coefficients, give a tiny contribution to the right-hand-side of the model.
Essentially, sparse regression fails to eliminate these spurious terms, since they give a very small contribution to the right-hand side of the equations and the overall error of the model. To remove spurious terms and get a sparser model, we restrict the library only to linear corrections in $\kappa$, at the expense of missing the $\kappa^2$ contribution to the coefficient of the fifth-order polynomial. In PySINDy, this is done by generating two libraries -- one at fifth order and a function of DMD coefficients $x$ and $y$, and another one at first order and function only of $\kappa$ -- and taking their tensor product. A model resulting from regression on this library is shown in~\eqref{eq:varkappa_nonnorm}. It was produced with a threshold of $\theta_{cr} = 10^{-3}$, corresponding to the minimum in the Pareto plot. The first line corresponds to the linear functions of $x$ and $y$, the second line to cubic terms, the third line to the fifth order terms, and the fourth line to the ``extra" spurious quadratic and fourth-order contributions. 
\begin{align}\label{eq:varkappa_nonnorm}
    x_t &= (0.073  + 4.38 \kappa) x + (25.37 + 2.64 \kappa)  y  \nonumber \\
&+ (-0.072 + 41.66 \kappa ) x^3 + (0.008 + 17.01 \kappa) x^2 y  + (- 0.073 + 39.73 \kappa ) x y^2   + (0.011 + 14.7 \kappa)y^3  \nonumber  \\
&+ (-0.001 -0.055 \kappa) x^5  - 0.27 x^4 y \kappa  - (0.001 + 0.45 \kappa)  x^3 y^2 - 0.5 x^2 y^3 \kappa -0.49 x y^4 \kappa -0.23 y^5 \kappa  +\\
&+ (0.022    + 0.043 x^2  - 0.041 x y - 0.14 y^2  -0.003 x^4 - 0.004 x^3 y  + 0.008 x^2 y^2 + 0.017 x y^3  + 0.017 y^4) \kappa \nonumber, \\
y_t &= (-25.37 -5.006 \kappa) x + (0.068  + 4.11 \kappa ) y \nonumber\\
&+(-0.011 -14.25  \kappa) x^3 + (-0.067  + 39.29 \kappa ) x^2 y  + (-0.011 - 13.82 \kappa ) x y^2 + (-0.069 + 40.38\kappa) y^3  \nonumber \\
&+ 0.2 x^5 \kappa + (- 0.002 + 0.069 \kappa ) x^4 y  + 0.4 x^3 y^2 \kappa + (- 0.003 + 0.17  \kappa) x^2 y^3 + 0.2 x y^4 \kappa  - (0.001 + 0.007 \kappa )  y^5 \nonumber \\
&+ (0.028   + 0.081 x^2 -0.016 x y -0.023 y^2  -0.005 x^4 + 0.002 x^3 y -0.005 x^2 y^2 -0.009 x y^3) \kappa  \nonumber.
\end{align}
The coefficients of this and the following models were rounded up to two significant digits for brevity. Linear and cubic terms are all proportional to $\kappa$. In addition, fifth-order nonlinearities feature $\kappa$ terms, missing their weaker constant counterparts. Note again, that all spurious terms are proportional to $\kappa$; hence SINDy again approximates the residual error of the solution with a combination of the smallest library terms. To lower their weight in the regression, we normalize $\kappa$ by the first value from our data set ($\kappa_{norm} = 0.0005$). This operation maps the values of the re-normalized parameter $\widetilde{\kappa}$ to $O(1)\div O(10)$, rescaling the coefficients proportional to $\kappa$ to very small values and thresholding them. This way, a sparser model is obtained for $\Theta_{cr} = 0.00023$, a minimum in the corresponding Pareto plot for the right-hand-side: 
\begin{align}\label{eq:varkappa_norm} 
    x_t &= (0.071 + 0.0012  \widetilde{\kappa}) x + (25.36  + 0.0019 \widetilde{\kappa}) y  \nonumber\\
&+(-0.07 + 0.02 \widetilde{\kappa} ) x^3  + (0.034 + 0.00025 \widetilde{\kappa} ) x^2 y  +  (- 0.07 + 0.02 \widetilde{\kappa}) x y^2  + (0.013  + 0.008 \widetilde{\kappa}) y^3 \nonumber \\
&-0.0014 x^5 -0.0027 x^3 y^2 -0.00074 x^2 y^3 -0.0014 x y^4 -0.00072 y^5, \\
y_t &= (-25.36  -0.00067  \widetilde{\kappa}) x + (0.069  + 0.0022 \widetilde{\kappa} )y  \nonumber\\
&+ (-0.014  - 0.0072 \widetilde{\kappa} ) x^3 + (-0.068 + 0.02 \widetilde{\kappa}) x^2 y + (- 0.014 - 0.007 \widetilde{\kappa} )  x y^2 + (-0.07 + 0.02  \widetilde{\kappa} ) y^3 \nonumber  \\
&+ 0.00059 x^5 -0.0013 x^4 y + 0.0012 x^3 y^2 -0.0026 x^2 y^3 + 0.00058 x y^4 -0.0013 y^5. \nonumber 
\end{align} 

The model contains additional linear terms in $\kappa$ that are not present in the original WNL model~\eqref{eq:wnl_H_fifth}. In contrast to the non-normalized previous model~\eqref{eq:varkappa_nonnorm}, this one recovers only the constant part of the fifth-order damping polynomial. At third order, however, all $\kappa$-terms balance the constant terms, as expected in the real part of $\chi$ of~\eqref{eq:wnl_chi35mu5}. Comparing with the WNL normal form~\eqref{eq:cart_norm_form}, the values of critical $\kappa_{cr}$ from the 8 cubic terms of the data-driven model~\eqref{eq:varkappa_norm} are where $\widetilde{\chi}_r = \widetilde{\chi}_{3,0} + \widetilde{\kappa} \widetilde{\chi}_{3, \widetilde{\kappa}}  = 0$, i.e.
\begin{eqnarray}
   \widetilde{\kappa}_{cr}  \in [3.426, 3.413, 3.449, 3.434]. \nonumber
\end{eqnarray}
The average value is compatible with $\widetilde{\kappa}_{cr} = \kappa_{cr}/\kappa_{norm} = 0.0017/0.005 = 3.4$ from the WNL analysis, turning supercritical bifurcation into subcritical. Thus, even without correction at the fifth order, the SINDy model is able to recover the subcritical behavior of the system. The amplitude of fifth order terms in this model is orders of magnitude smaller than in the model~\eqref{eq:sub_cart_varD}, again likely due to the different normalization of $\kappa$ and the DMD coefficients. 

\subsection{Full model depending on both $D$ and $\kappa$}
\begin{figure}
\centering
     \begin{subfigure}[b]{0.47\textwidth}
         \centering
         \includegraphics[width=\textwidth]{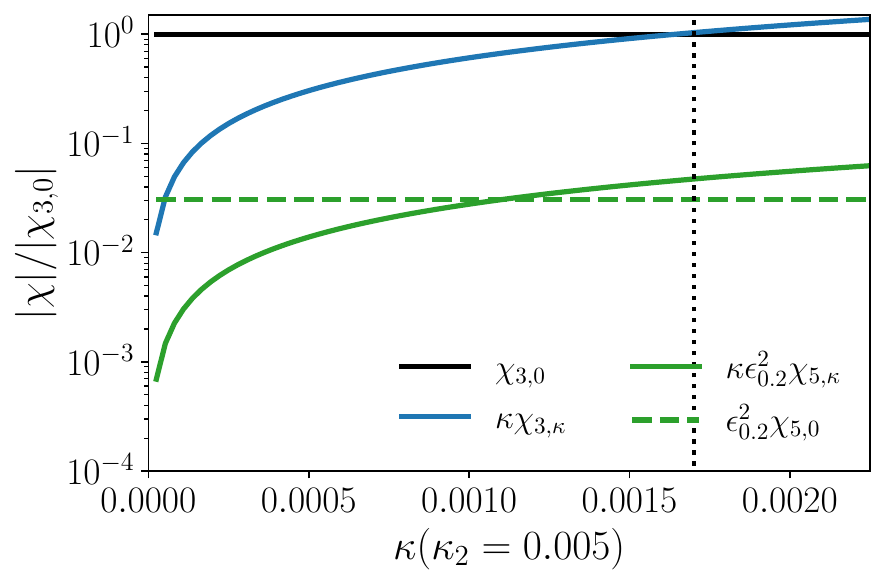}
          \caption{}
     \end{subfigure}
     \hfill 
          \begin{subfigure}[b]{0.47\textwidth}
         \centering
         \includegraphics[width=\textwidth]{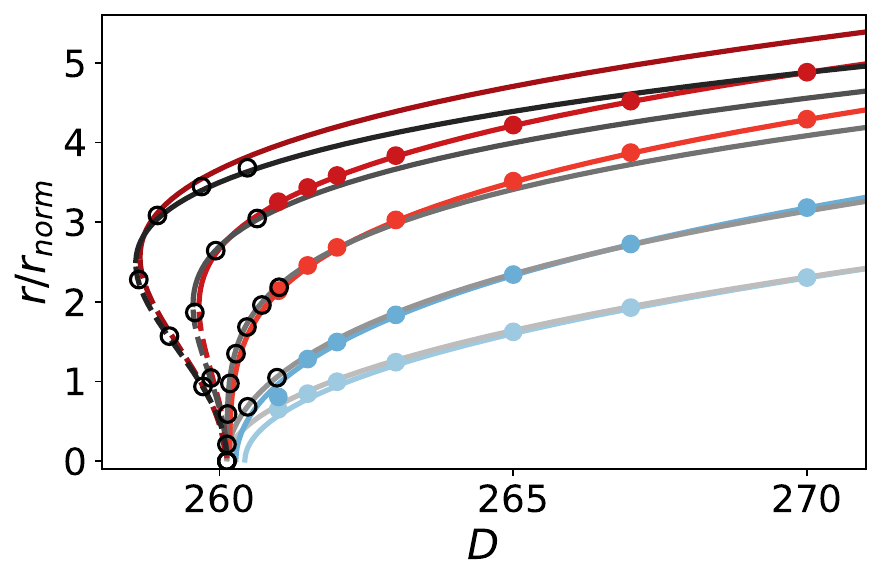}
          \caption{}
     \end{subfigure}
     
    \caption{(a) Contributions to $\chi$, as a function of $\kappa$ at third order. Vertical dotted line denotes $\kappa=0.0017$, where the dynamo becomes subcritical. The data taken at $D=270$ ($\epsilon=0.2$, the highest value in the training range). 
    (b) Bifurcation diagram of the large full dynamical model of the system as a function of $D$ and $\kappa$. Black solid lines, WNL model~\eqref{eq:wnl_H_fifth}, in colour, SINDy model~\eqref{eq:sindy_varkappa_varD}. Filled circles, amplitudes of the DMD coefficients; empty circles, solution branches obtained with Newton-Krylov method. In blue, supercritical bifurcation, in red, subcritical bifurcation. Increasing color intencity: $\kappa_1 = [0,0.2,0.34,0.4, 0.44]$, $\kappa_2 = 0.005$. Dashed lines, unstable dynamo branches.}
    \label{fig:SINDy_varkappa_varD}
\end{figure}
Equipped with findings from sections~\ref{sec:varD} and~\ref{sec:varkappa_fixedD}, we finally develop a model that traces both transitions to dynamo waves with $D$ and from supercriticality to subcriticality with $\kappa$. As previously, the aim is to obtain a model as close as possible to the normal form and avoid spurious terms, so we restrict parameter dependence of the library to multiples of $D$ and $\kappa$ with powers of $x$ and $y$, omitting terms like $\kappa^2$ and $\kappa D$. We span values of $D \in [240, 270]$,  $\kappa_2 = 5 \cdot 10^{-3}$,  and $\kappa = \kappa_1 \kappa_2 \in [0, 2.25 \cdot 10^{-3}]$, with $111$ different runs with parameter pairs $(D, \kappa)$ in total. Using re-normalization as in section~\ref{sec:varkappa_fixedD}, a threshold of $\Theta_{cr} = 0.0001$ and finite difference differentiation, we find 
\begin{align}\label{eq:sindy_varkappa_varD}
    x_t &= (-10.3 + 0.04 D + 0.0032 \widetilde{\kappa}  ) x  + (8.75 + 0.063 D + 0.00037 \widetilde{\kappa} ) y \nonumber \\
    &+ (- 0.2 + 0.0005 D + 0.02  \widetilde{\kappa} ) x^3 + (- 0.019  + 0.00013 D + 0.0074 \widetilde{\kappa}  ) x^2 y  \nonumber \\
    & + (-0.21  + 0.00052 D + 0.02 \widetilde{\kappa} ) x y^2 + (-0.019 + 0.00013 D + 0.0073 \widetilde{\kappa}  ) y^3  \nonumber \\
    &-0.0013 x^5 -0.00064 x^4 y -0.0027 x^3 y^2 -0.0013 x^2 y^3  -0.0013 x y^4 - 0.00064 y^5, \\
y_t & = (-8.8  - 0.063 D - 0.00029  \widetilde{\kappa} ) x  + (-10.3 + 0.04 D + 0.0029 \widetilde{\kappa}  ) y \nonumber \\
&+ ( 0.027 -0.00016 D -0.0074  \widetilde{\kappa} ) x^3 + (- 0.2 + 0.00052 D + 0.02 \widetilde{\kappa}  ) x^2 y \nonumber \\ 
& + (0.032 - 0.00017 D -0.0075 \widetilde{\kappa}  ) x y^2 + ( -0.2 + 0.00051 D + 0.02 \widetilde{\kappa} ) y^3   \nonumber \\
&+ 0.00063 x^5 -0.0013 x^4 y + 0.0013 x^3 y^2 -0.0027 x^2 y^3 + 0.00064 x y^4 -0.0013 y^5 \nonumber.
\end{align}
This model has a similar dependence on $D$ as the model~\eqref{eq:sub_cart_varD}, and captures the linear part of the Hopf bifurcation. It also recovers the competition between constant and proportional to $\kappa$ coefficients for the third-order polynomial terms. In addition, it identifies a weak dependence of the cubic term on $\epsilon^2$, and thus $D$, at the third order, as is indeed the case in the WNL form~\eqref{eq:wnl_H_fifth}. The comparable term $\kappa \epsilon^2 \chi_{5,\kappa}$  (figure~\ref{fig:SINDy_varkappa_varD}a) was not identified since the corresponding interaction did not enter the function library. The model is quantitatively consistent with the $\kappa$-only model~\eqref{eq:varkappa_norm},  when $D=262$ is plugged in into the linear and cubic terms.  As in~\eqref{eq:varkappa_norm}, the fifth order polynomials contain only constant terms and the dependence on $\kappa$ is not recovered at fifth order. Alternative models, without re-normalization of $\kappa$, or restriction on the library terms, penalize the constant part of the third-order and fifth-order coefficients,  and again promote spurious correlations between $x^2$, $y^2$, $x^4$, $y^4$, and $\kappa$, similarly to~\eqref{eq:varkappa_nonnorm} in section~\ref{sec:varkappa_fixedD}.  On the contrary, model~\eqref{eq:sindy_varkappa_varD} recovers both the dependence on $D$ and the competition between dissipative and $\kappa$-effects at third order, hence transitioning from supercriticality to subcriticality. It is thus a sufficient representation of the WNL normal form~\eqref{eq:wnl_H_fifth} of the dynamo cycle.

\section{Discussion}\label{sec:discussion}

While the reconstruction of the third-order supercritical normal form is robust both at the onset and away from it, reconstruction of fifth-order model is challenging and requires special treatment of signals -- filtering or Hankel DMD with high delay parameter. Even then, equations~\eqref{eq:sub_wnl5}, constructed from several runs at one point in the parameter space at onset, contain spurious fourth order terms. These act as additional damping terms and equally well approximate the flow dynamics with weakened the fifth-order term (figure~\ref{fig:wnl_sindy}d). It is likely that the dynamical information contained in the polar representation of the normal form is not sufficient to identify these terms using SINDy, as similar powers of $r$ can give similar contributions to short algebraically growing signals like those of figure~\ref{fig:wnl_sindy}(a).

On the other hand, a correct subcritical fifth order model can be obtained with ``Cartesian" representation of the data, after normalization and filtering. Equations~\eqref{eq:sub_cart_varD} contain all the terms from the fifth order normal form, depending on the governing parameter of the system, dynamo number $D$. Such equations allow us to extrapolate the results of both subcritical and supercritical normal forms away from the bifurcation point, to values of $\epsilon^2$ as large as $D^c$ itself. The accuracy of the model is better for the subcritical fifth-order model, as it saturates faster (figure~\ref{fig:model_varD}d). Both SINDy and WNL third-order models for supercritical bifurcation systematically overestimate magnetic field amplitude away from the bifurcation point (figure~\ref{fig:model_varD}b). Both models have all the desired properties of the system: one stable fixed point at $r=0$ exists below $D^c \approx 260$, which later becomes unstable, and the limit cycle appears.

Model~\eqref{eq:sindy_varkappa_varD}, captures both $D$ and $\kappa$ dependence, and represents the transitions between supercritical and subcritical dynamo states of the equivalent WNL dynamo form (figure~\ref{fig:SINDy_varkappa_varD}b). It also learned the second most important term in~\eqref{eq:wnl_H_fifth}, $\epsilon^2 \chi_{5,0}$ in the third-order polynomial nonlinearity, which caused overestimation of the dynamo magnitude in figure~\ref{fig:model_varD}(b). Still, its library was constrained only to the most essential terms, and the model~\eqref{eq:sindy_varkappa_varD} misses a part of the dynamics: the terms proportional to $\kappa D$ at third order, and the terms proportional to $\kappa$ at fifth order. In addition, the weak linear dependence on $\kappa$ of the linear terms leads to weak changes in the critical value of the dynamo number $D_{cr}$ for the onset of the unstable branch, as shown in figure~\ref{fig:SINDy_varkappa_varD}(b).  Nevertheless, it captures both stable branches of the dynamo (solid lines in figure~\ref{fig:SINDy_varkappa_varD}), on which it has been trained, and unstable branches (in dashed in figure~\ref{fig:SINDy_varkappa_varD}). The solution of the dynamo equations~\eqref{eq:dynamo} does not approach these branches in our data, and therefore SINDy regression discovered the parts of the dynamics that it has not seen before. To cross-check the extent of unstable branches near the onset of the dynamo instability, we perform an exact Newton--Krylov tracking of the dynamo fixed points, following the method described by~\citet{Chandler_Kerswell_2013} and the code developed by~\citet{Skene_Marcotte_Tobias_2025} which performs this using Dedalus. This code leverages the Newton--Krylov solver developed by~\citet{Willis_2017} which is publicly available here~\citep{Willis_NH}. The amplitudes of the magnetic field of these exact solutions, added to figure~\ref{fig:SINDy_varkappa_varD}b, are in a good agreement with both the WNL analysis and SINDy results at the onset of dynamo instability. However, the accuracy of SINDy models is better than the WNL ones with respect to the actual dynamo data (DMD amplitudes, filled circles in figure~\ref{fig:SINDy_varkappa_varD}b) for increasing levels of subcriticality and $D$.  Despite its shortcomings, the SINDy model approximate the full dynamics of the system sufficiently well away from the dynamo onset, and has a good capacity for extrapolation.

Throughout, we have found that it is challenging to identify the correct normal forms without a preliminary knowledge of the target equations for our dynamo system. We found that the indicators giving confidence in a model are (i) a robust minimum in Pareto plot, i.e. the resultant model does not change when the cut-off threshold is slightly modified; (ii) the insensitivity of results to different methods of differentiation (e.g. finite or smoothed finite differences); (iii) the robustness of the model with respect to perturbation of initial conditions and extrapolation to different values of $D$ outside of the region where the data were collected. We suggest that the difficulties in identifying reduced dynamo equations comes from ``stiffness" of the saturation mechanisms in the dynamo equations~\eqref{eq:dynamo}. Indeed, WNL theory assumes that nonlinearities can be expanded in Taylor series as a weak perturbation around the system state. However, for nonlinear terms like $1/(1+\kappa_2 B^2)$ in equations~\eqref{eq:dynamo}, this expansion is strictly valid only for $|B|<1/\sqrt{\kappa_2}$. Thus, the WNL form~\eqref{eq:wnl_H_fifth} has systematic error with respect to the data for increasing values of $\kappa_2$ and should be expected to fail completely for large values. 

For such a non-analytic nonlinearity, WNL theory has a limited radius of convergence in parameters, independent of epsilon. In this work, we chose a particular value of $\kappa_2 = 5 \cdot 10^{-3}$, where we found the condition  $|B|<1/\sqrt{\kappa_2}\sim  14$ to be empirically satisfied (figure~\ref{fig:solutions_dmd}a). SINDy, on the other hand, has no such requirements on the amplitude of $B$, and can learn models representing the normal form in regimes inaccessible to WNL analysis. We leave developing models with stiff subcriticalities, as well as more detailed representation of nonlinear dynamics at the fifth order for the future work, exploring more robust methods of SINDy regression. We expect our models to be consistent with the normal forms of~\citet{tobias1995chaotically}, describing the dynamo as a saddle node/Hopf bifurcation of two degrees of freedom in the magnetic field and one in the velocity perturbation, especially close to the bifurcation point where the velocity is enslaved to the  magnetic field components. Furthermore, our reconstructed model highlights the importance of cubic nonlinearities, recently discovered in empirical data-driven models of the solar cycle~\citep{bonanno2025data}.

When $D$ is much larger than $D^c$ and DMD is performed on steady-state data,  larger deviations from eigenvectors arise, especially in $\Psi(B_G)$ which is directly affected by the nonlinearity~\eqref{eq:dynamo}. These deviations reflect the contribution of high-order triadic modal interactions to the dominant frequency in the flow, $\mathrm{i} \omega_i^c t$, and the corresponding spatial structure. For some cases, our preliminary results show that ``averaging" the eigenvectors allows us to derive more robust dynamo equations, extrapolable to high values of $D$, at the expense of reduced accuracy at the dynamo onset. Understanding this effect, and exploring the potential of SINDy to capture secondary Hopf bifuractions to more complex, quasi-periodic dynamo solutions, are other promising directions for future work.

\section{Conclusions}\label{sec:conclusion}

In this work, we implemented a robust framework to discover supercritical and subcritical reduced equations for dynamo cycles, using DMD and SINDy, together with an approach to estimate DMD model coefficients through the adjoint DMD operator, that, to our knowledge, has not been used before. We found that third-order models -- obtained either from SINDy or WNL analysis -- underestimate the dynamo magnitude by up to $35\%$ away from the bifurcation point. Subcritical fifth-order models are much more difficult to obtain than supercritical ones, due to the stronger impact of nonlinearity on the dynamo behavior, and much larger polynomial library needed for regression. Nevertheless, after filtering quadratic nonlinear contributions to the eigenvector coefficients, and constraining the library to essential and normalized contributions of $\kappa$, we were able to construct a model that recovers both the transition to the dynamo cycle from steady, non-magnetic solutions, and infer the parametric separation between supercritical and subcritical regimes. Our SINDy-produced reduced order model reproduces both the stable and unstable dynamo branches, having only been trained on data showing growth to the stable branch. This means that SINDy can successfully infer the unstable branch by learning the correct terms in the normal form, bypassing the need to carefully construct training data on the elusive unstable orbit. Furthermore, the models are robust and integrable up to ten times further away from the training parameter regions, and extrapolate well away from the onset, showing their potential for approximating magnetic field strength for extreme planetary and stellar parameters. In the future work, we will expand this analysis to more complex, three-dimensional, dynamo cycles, such as those observed in models of stellar convective envelopes.

\bibliographystyle{IEEEtranN}
\bibliography{dynbib} 

\appendix

\section{Third-order expansion}\label{sec:a_wnl_terms_3}

To third order in $\epsilon$ we get
\begin{equation}\label{eq:AppA_exp}
\textbf{q}=
(A,B)^T
=
\left[
\epsilon G \textbf{q}_G \exp(\mathrm{i}\omega_i^c t)
+ \epsilon^3 \textbf{q}_{3,\omega_i^c} \exp(\mathrm{i}\omega_i^c t) +\textbf{q}_{3,3\omega_i^c} \exp(3\mathrm{i}\omega_i^c t) 
+\textrm{c.c}\right] + \mathcal{O}(\epsilon^5),
\end{equation}
and
\begin{equation}\label{eq:expansion_u}
u = \epsilon^2\left[(G^2u_{G^2}\exp(2\mathrm{i}\omega_i^ct) + \textrm{c.c}) + |G|^2u_{|G|^2} \right] + \mathcal{O}(\epsilon^4).
\end{equation}
Here, $\textbf{q}_G$ is the eigenvector that solves (\ref{eq:eigenvalue_problem_AB}) at $D=D^c$, and $G(t_2)$ is its amplitude as a function of the slow time $t_2 = \epsilon^2 t$. For values close to criticality, as measured by $\epsilon$, we get additional terms stemming from nonlinear terms of~\eqref{eq:dynamo}. These include $u_{G^2}$ which is a velocity correction with twice the frequency of the original dynamo instability, and base flow modification term $u_{|G|^2}$. To more concisely write the equations we now define
\begin{gather}
H_u(A_\cdot,B_\cdot) =  C_3A_\cdot\partial_x(B_\cdot) - C_4B_\cdot\partial_x(A_\cdot), \\
H_B(A_\cdot,u_\cdot) = C_1 D^c \partial_x(u_\cdot)A_\cdot - C_2 D^c u_\cdot (A_\cdot)_x, \\
    T(q_1, q_2, q_3) = -((q_1)_{xx} + C_\eta q_1)q_2q_3,
\end{gather}
where here and afterwards we write $A_\cdot$ and $B_\cdot$ for the $A$ and $B$ components of $\mathbf{q}_\cdot$, respectively.
The equations for these terms are
\begin{align}
(2\mathrm{i}\omega_i^c - \mathcal{L}_u)u_{G^2} &= H_u(A_G, B_G), \label{eq:resolvent1}\\
- \mathcal{L}_u u_{|G|^2} &= H_u(A_G, \overline{B_G}) + \textrm{c.c}.\label{eq:resolvent2}
\end{align}
This shows that these higher order terms stem from nonlinearities of the first order instability with itself. In general, the base flow is also modified due to increase of the critical parameter, however in our case the base flow is zero and no such parameter-induced modification takes place. 

At third order we then get corrections to the magnetic terms stemming from interactions of the instability at first order, with velocity corrections at second order. Again, from \eqref{eq:dynamo} and solutions of equations \eqref{eq:resolvent1} and \eqref{eq:resolvent2} we obtain
\begin{equation}
(\mathrm{i}\omega_i^c - \mathcal{L}_{A,B})\mathbf{q}_{3,\omega_i^c} = -\mathbf{q}_G\frac{\partial G}{\partial t_2} + G\boldsymbol{\eta}_3 - |G|^2G\boldsymbol{\chi}_3,
\end{equation}
with 
\begin{equation}
\boldsymbol{\eta}_3 = 
(
0,
-D^c C_\Omega \sin(\pi x/2){A_G}_x)^T, \qquad \boldsymbol{\chi}_3 = -\left(\boldsymbol{\chi}_{u} + \kappa_1\kappa_2\boldsymbol{\chi}_{\kappa}\right),
\end{equation}
where
\begin{equation}
    \boldsymbol{\chi}_{u}=
\begin{pmatrix}
0 \\
H_B(\overline{A_G}, u_{G^2}) + H_B(A_G, u_{|G|^2}) 
\end{pmatrix},
\end{equation}
\begin{equation}
    \boldsymbol{\chi}_{\kappa}=
\begin{pmatrix}
-2T(A_G, B_G, \overline{B_G})-T(\overline{A_G}, B_G, B_G)\\
-2T(B_G, B_G, \overline{B_G})-T(\overline{B_G}, B_G, B_G)
\end{pmatrix}.
\end{equation}
This equation is singular, since $\mathrm{i} \omega_i^c$ is chosen to be an eigenvalue of $\mathcal{L}_{A,B}$. Therefore, in order for our expansion to be valid we require - by Fredholm's theorem - the right hand side to be orthogonal to the adjoint eigenfunctions spanning the nullspace of operator $(-\mathrm{i}\omega_i^c - \mathcal{L}_{A,B}^\dagger)$, where $\dagger$ denotes the adjoint operator of $\mathcal{L}_{A,B}$.   $\mathcal{L}_{A,B}^\dagger$ is chosen such that $\langle \mathbf{p}^\dagger, \mathcal{L}_{A,B} \mathbf{p}\rangle = \langle \mathcal{L}_{A,B}^\dagger\mathbf{p}^\dagger, \mathbf{p}\rangle$ with inner product
\begin{equation}
    \langle \mathbf{a}, \mathbf{b}\rangle = \int_0^L  \overline{\mathbf{a}} \cdot \mathbf{b}\;\textrm{d}x.
\end{equation}
for all $\mathbf{p}^\dagger$ and $\mathbf{p}$. Integration by parts yields
\begin{equation}
\mathcal{L}_{A,B}^\dagger = 
\begin{pmatrix}
    \partial_{xx} + C_\eta & C_\Omega D^c \pi /2 \cos(\pi x/2) + C_\Omega D^c \sin(\pi x/2)  \partial_x \\
    C_\alpha \cos(\pi x/2) & \partial_{xx} + C_\eta
\end{pmatrix}.
\end{equation}
Thus, by solving the adjoint eigenvalue problem $(-\mathrm{i}\omega_i^c-\mathcal{L}_{A,B}^\dagger)\mathbf{q}^\dagger_G=0$, normalising such that $\langle \mathbf{q}^\dagger_G, \mathbf{q}_G\rangle=1$, and applying the solvability condition we obtain the amplitude equation
\begin{equation}\label{eq:STLandau}
\frac{\partial G}{\partial t_2} = \eta_3 G - \chi_3 |G|^2G.
\end{equation}

\section{Fifth-order expansion}\label{sec:a_wnl_terms_5}
To proceed to fifth order, we now introduce the slower timescale $t_4$, where $t_4=\epsilon^4 t$, and write that the amplitude $G(t_2, t_4)$ evolves on these two independent timescales. This naturally implies that
$
\frac{\textrm{dG}}{\textrm{d}t} = \epsilon^2\partial_{t_2}G + \epsilon^4\partial_{t_4}G.
$
Our third order expansion has already determined $\partial_{t_2}G$. To find $\partial_{t_4}G$ we now extend our analysis to fifth order. To this end, we start by finding the explicit form of terms at third order. These are 
\begin{gather}
(3\mathrm{i}\omega_i^c - \mathcal{L}_{A,B})\mathbf{q}_{G^3}= (0, H_B(A_G, u_{G^2}))^T, \\
(\mathrm{i}\omega_i^c - \mathcal{L}_{A,B})\mathbf{q}_{|G|^2G}= \chi_u\mathbf{q}_G + \boldsymbol{\chi}_{u},
\label{equ:third_G}
\\
(\mathrm{i}\omega_i^c - \mathcal{L}_{A,B})\mathbf{q}_{G,3} = -\eta_3\mathbf{q}_G + \boldsymbol{\eta}_3,
\label{equ:third_G3}
\\
(3\mathrm{i}\omega_i^c - \mathcal{L}_{A,B})\mathbf{q}_{G^3,\kappa}= (T(A_G,B_G,B_G), T(B_G,B_G,B_G))^T,\\
(\mathrm{i}\omega_i^c - \mathcal{L}_{A,B})\mathbf{q}_{|G|^2G,\kappa}= \chi_\kappa \mathbf{q}_G + \boldsymbol{\chi}_\kappa. 
\end{gather}

These terms constitute the full third order term in the expansion as
\begin{equation}\label{eq:AppB_exp3}
\mathbf{q}_3 = G^3\left(\mathbf{q}_{G^3}+\kappa\mathbf{q}_{G^3,\kappa} \right)\exp(3\mathrm{i}\omega_i^ct) + \left(|G|^2G\mathbf{q}_{|G|^2G} +\kappa|G|^2G\mathbf{q}_{|G|^2G, \kappa} + G\mathbf{q}_{G,3}  \right)\exp(\mathrm{i}\omega_i^ct) + \textrm{c.c}.
\end{equation}
We note that equations (\ref{equ:third_G}) and (\ref{equ:third_G3}) are solvable precisely because of our third order amplitude equation, and therefore have a unique solution that is orthogonal to $\mathbf{q}_G^\dagger$ - which is the solution we take for our expansion. 

With these terms found we can proceed to fourth order, where we obtain
\begin{gather}
(2\mathrm{i}\omega_i^c-\mathcal{L}_u)u_{G^2|G|^2} =  
H_u(\overline{A}_G, B_{G^3}) + 
H_u(A_{G^3}, \overline{B_{G}}) +
H_u(A_G, B_{|G|^2G}) +
H_u(A_{|G|^2G}, B_{G}) ,\\
(4\mathrm{i}\omega_i^c-\mathcal{L}_u)u_{G^4} = 
H_u(A_G, B_{G^3}) + H_u(A_{G^3}, B_G) ,\\
-\mathcal{L}_u u_{|G|^2,4} = 
H_u(\overline{A_G}, B_{G,3}) +
H_u(\overline{A_{G,3}}, B_G) + \textrm{c.c} ,\\
-\mathcal{L}_u u_{|G|^4} = 
H_u(\overline{A_G}, B_{|G|^2G}) +
H_u(A_{|G|^2G}, \overline{B_{G}})
+ \textrm{c.c.} ,\\
-\mathcal{L}_u u_{|G|^4,\kappa} = 
H_u(\overline{A_G}, B_{|G|^2G, \kappa}) +
H_u(A_{|G|^2G, \kappa}, \overline{B_{G}})
+ \textrm{c.c.} ,\\
(2\mathrm{i}\omega_i^c-\mathcal{L}_u) u_{G^2,4} = 
H_u(A_G, B_{G,3}) + 
H_u(A_{G,3}, B_{G}) ,\\
(2\mathrm{i}\omega_i^c-\mathcal{L}_u)u_{G^2|G|^2,\kappa} =
H_u(\overline{A}_G, B_{G^3}) + 
H_u(\overline{A}_{G^3}, \overline{B_{G}}) +
H_u(A_G, B_{|G|^2G}) +
H_u(A_{|G|^2G}, B_{G}) ,\\
(4\mathrm{i}\omega_i^c-\mathcal{L}_u)u_{G^4,\kappa} = 
H_u(A_G, B_{G^3,\kappa}) + H_u(A_{G^3, \kappa}, B_{G}).
\end{gather}
These terms give the complete fourth order description as
\begin{gather}
u_4 = \left[(G^2|G|^2u_{G^2|G|^2} + G^2u_{G^2,4} + \kappa G^2|G|^2u_{G^2|G|^2, \kappa} )\exp(2\mathrm{i}\omega_i^c t) + G^4(u_{G^4} + \kappa u_{G^4, \kappa} )\exp(4\mathrm{i}\omega_i^c t) + \textrm{c.c}\right] +\nonumber\\
|G|^2u_{|G|^2,4} + |G|^4(u_{|G|^4} + \kappa u_{|G|^4, \kappa}).
\end{gather}

Similarly to third order, at fifth order we now obtain an amplitude equation by applying a solvability condition to singular fifth order equations. This equation determines the dependence of the amplitude $G$ on the slower timescale $t_4$
\begin{equation}
\frac{\partial G}{\partial t_4} = \eta_5 G - \chi_5 |G|^2G + \mu_5|G|^4G.
\end{equation}
The coefficients to this equation are given by
\begin{equation}
    \eta_5 = \langle B^\dagger_G, -C_\Omega \sin(\pi x/2) D^c A_{G,3,x} \rangle,
\end{equation}
\begin{gather}
    \chi_5 = -\langle B^\dagger_G, H_B(\overline{A_G}, u_{G^2,4}) + H_B(A_G, u_{|G|^2,4}) + H_B(A_{G,3}, u_{|G|^2}) + H_B(\overline{A_{G,3}}, u_{G^2}) + H_B(A_{G}, u_{|G|^2}) +\nonumber\\
    H_B(\overline{A_{G}}, u_{G^2}) - C_\Omega \sin(\pi x/2) D^c A_{|G|^2G,3,x} \rangle,
\end{gather}
\begin{gather}
    \mu_5 = \langle B^\dagger_G, H_B(A_G, u_{|G|^4}) +
    H_B(\overline{A_G}, u_{G^2|G|^2}) + 
    H_B(A_{G^3}, \overline{u_{G^2}}) + \nonumber\\
    H_B(A_{|G|^2G}, u_{|G|^2}) + 
    H_B(\overline{A_{|G|^2G}}, u_{G^2})
    \rangle.
\end{gather}
The terms involving $\kappa$ are
\begin{gather}
    \mu_{5,\kappa} = \langle B^\dagger_G, H_B(A_G, u_{|G|^4, \kappa}) +
    H_B(\overline{A_G}, u_{G^2|G|^2, \kappa}) + 
    H_B(A_{G^3, \kappa}, \overline{u_{G^2}}) +
    H_B(A_{|G|^2G, \kappa}, u_{|G|^2}) + \nonumber\\
    H_B(\overline{A_{|G|^2G, \kappa}}, u_{G^2}) +
    T(B_{G^3},\overline{B_G},\overline{B_G}) +
    2T(\overline{B_{G}},B_{G^3},\overline{B_G}) + 
    2T(B_{|G|^2G},B_{G},\overline{B_G}) + 
    2T(B_{G},B_{|G|^2G},\overline{B_G}) + \nonumber\\
    T(\overline{B_{|G|^2G}},B_{G},B_G) + 
    2T(B_{G},\overline{B_{|G|^2G}},B_G) + 
    T(\overline{B_{G}},B_{|G|^2G},B_G)
    \rangle+
    \nonumber\\
    \langle A^\dagger_G,
    T(A_{G^3},\overline{B_G},\overline{B_G}) +
    2T(\overline{A_{G}},B_{G^3},\overline{B_G}) + 
    2T(A_{|G|^2G},B_{G},\overline{B_G}) + 
    2T(A_{G},B_{|G|^2G},\overline{B_G}) + \nonumber\\
    T(\overline{A_{|G|^2G}},B_{G},B_G) + 
    2T(A_{G},\overline{B_{|G|^2G}},B_G) + 
    T(\overline{A_{G}},B_{|G|^2G},B_G)
    \rangle,
\end{gather}
\begin{gather}
    \mu_{5,\kappa^2} = \langle B^\dagger_G,
    T(B_{G^3,\kappa},\overline{B_G},\overline{B_G}) +
    2T(\overline{B_{G}},B_{G^3,\kappa},\overline{B_G}) + 
    2T(B_{|G|^2G,\kappa},B_{G},\overline{B_G}) + 
    2T(B_{G},B_{|G|^2G,\kappa},\overline{B_G}) + \nonumber\\
    T(\overline{B_{|G|^2G,\kappa}},B_{G},B_G) + 
    2T(B_{G},\overline{B_{|G|^2G,\kappa}},B_G) + 
    T(\overline{B_{G}},B_{|G|^2G,\kappa},B_G)
    \rangle + \nonumber\\
    \langle A^\dagger_G,
    T(A_{G^3,\kappa},\overline{B_G},\overline{B_G}) +
    2T(\overline{A_{G}},B_{G^3},\overline{B_G}) + 
    2T(A_{|G|^2G,\kappa},B_{G},\overline{B_G}) + 
    2T(A_{G},B_{|G|^2G,\kappa},\overline{B_G}) + \nonumber\\
    T(\overline{A_{|G|^2G,\kappa}},B_{G},B_G) + 
    2T(A_{G},\overline{B_{|G|^2G,\kappa}},B_G) + 
    T(\overline{A_{G}},B_{|G|^2G,\kappa},B_G)
    \rangle,
\end{gather}
\begin{gather}
    \mu_{5,\kappa \kappa_2} = -\langle B^\dagger_G,
    6T(B_G, B_G^2, \overline{B_G}^2)
    + 4T(\overline{B_G}, B_G^3, \overline{B_G})
    \rangle - \nonumber\\
    \langle A^\dagger_G,
    6T(A_G, B_G^2, \overline{B_G}^2)
    + 4T(\overline{A_G}, B_G^3, \overline{B_G})
    \rangle,
\end{gather}
and
\begin{gather}
    \chi_{5,\kappa} = \langle B^\dagger_G, 
    2T(B_{G,3}, B_G, \overline{B_G}) +
    T(\overline{B_{G,3}}, B_G, B_G) +
    2T(B_{G}, B_{G,3}, \overline{B_G}) + \nonumber\\
    T(B_{G}, \overline{B_{G,3}}, B_G) + 
    2T(\overline{B_{G}}, B_{G,3}, B_G) - 
    C_\Omega \sin(\pi x/2)D^c A_{G,3,\kappa,x}
    \rangle +\nonumber\\
    \langle A^\dagger_G, 
    2T(A_{G,3}, B_G, \overline{B_G}) +
    T(\overline{A_{G,3}}, B_G, B_G) +
    2T(A_{G}, B_{G,3}, \overline{B_G}) + \nonumber\\
    T(A_{G}, \overline{B_{G,3}}, B_G) + 
    2T(\overline{A_{G}}, B_{G,3}, B_G) \rangle.
\end{gather}

\end{document}